\documentclass[fleqn,usenatbib]{mnras}

\usepackage{newtxtext,newtxmath}

\usepackage[T1]{fontenc}
\usepackage{ae,aecompl}

\usepackage{graphicx}	
\usepackage{amsmath}	
\usepackage{amssymb}	
\usepackage{ulem}	

\title[Observing and modelling EK Draconis]{Observing and modelling the young solar analogue EK Draconis: starspot distribution, elemental abundances, and evolutionary status}
\author[H.V. \c{S}enavc{\i} et al.]{
H.V. \c{S}enavc{\i}$^{1}$\thanks{E-mail: hvsenavci@ankara.edu.tr},
T. K{\i}l{\i}\c{c}o\u{g}lu$^{1}$,
E. I\c{s}{\i}k$^{2,3}$,
G.A.J. Hussain$^{4,5}$,
D. Montes$^{6}$,
E. Bahar$^{1}$ and
\newauthor
S.K. Solanki$^{3,7}$
\\
$^{1}$Department of Astronomy and Space Sciences, Faculty of Science, Ankara University, Tando\u{g}an, Ankara, Turkey\\
$^{2}$Department of Computer Science, Turkish-German University, \c{S}ahinkaya Cd. 108, 34820 Beykoz, Istanbul, Turkey \\
$^{3}$Max-Planck-Institut f\"{u}r Sonnensystemforschung, Justus-von-Liebig-Weg 3, 37077, G\"{o}ttingen, Germany\\
$^{4}$European Southern Observatory, Karl-Schwarzschild-Str. 2, 85748 Garching bei M\"unchen, Germany\\
$^{5}$CNRS/IRAP, Universit\'e de Toulouse, 14 Avenue E. Belin, 31400, Toulouse, France\\
$^{6}$Departamento de F{\'i}sica de la Tierra y Astrof{\'i}sica \& IPARCOS-UCM, 
Facultad de Ciencias F{\'i}sicas, \\ Universidad Complutense de Madrid, E-28040 Madrid, Spain \\
$^{7}$School of Space Research, Kyung Hee University, Yongin 446-701, Gyeonggi, Korea
}

\date{Accepted 2021 January 13.}

\pubyear{2020}

\begin{document}
\label{firstpage}
\pagerange{\pageref{firstpage}--\pageref{lastpage}}
\maketitle
\definecolor{forestgreen}{rgb}{0.00, 0.30, 0.00}
\definecolor{purple}{rgb}{0.50, 0.0, 0.50}

\begin{abstract}

Observations and modelling of stars with near-solar masses in their early phases of evolution are 
critical for a better understanding of how dynamos of solar-type stars evolve. 
We examine the chemical composition and the spot distribution of the pre-main-sequence solar analogue 
EK Dra. Using spectra from the HERMES Spectrograph (La Palma), we obtain the abundances 
of 23 elements with respect to the solar ones, which lead to a $[{\rm Fe/H}]=0.03$, with significant overabundance of 
Li and Ba. The s-process elements Sr, Y, and Ce are marginally overabundant, while Co, Ni, Cu, Zn 
are marginally deficient compared to solar abundances. The overabundance of Ba is most likely due to the assumption of depth-independent microturbulent velocity. Li abundance is consistent with the age and the other 
abundances may indicate distinct initial conditions of the pre-stellar nebula. We estimate 
a mass of 1.04 $M_\odot$ and an age of $27^{+11}_{-8}$\,Myr using various spectroscopic and photometric indicators. 
We study the surface distribution of dark spots, using 17 spectra collected during 15 nights 
using the CAFE Spectrograph (Calar Alto). 
We also conduct flux emergence and transport (FEAT) simulations for EK Dra's parameters and 
produce 15-day-averaged synoptic maps of the likely starspot distributions. 
Using Doppler imaging, we reconstruct the surface brightness distributions for the observed 
spectra and FEAT simulations, which show overall agreement for polar 
and mid-latitude spots, while in the simulations there is a lack of low-latitude spots compared to the observed image. We find indications that cross-equatorial extensions of mid-latitude spots 
can be artefacts of the less visible southern-hemisphere activity. 


\end{abstract}




\section{Introduction}
\label{sec:intro}


Monitoring young solar analogues is of key importance in determining
patterns of magnetic activity on the Sun during its early stages of 
evolution \citep[e.g.][]{Kriskovics2019}. Such observations are essential for testing theoretical 
models of the physical processes underlying stellar magnetic activity, namely, 
the generation, emergence, and surface transport of magnetic flux in solar-type stars. A fundamental problem awaiting explanation is the surface 
distribution of spots on such young and active stars, which is very different from 
solar patterns. 
In addition, investigating magnetic activity in young solar analogues is of great importance in understanding the angular momentum evolution \citep{Cang2020} and the stellar dynamo, as they strongly affect each other throughout stellar evolution \citep[][]{gudel07,brunbrowning17}. Probing stellar magnetic activity also holds special interest, since magnetic activity has important effects on the structure and evolution of planetary atmospheres \citep{johnstone17}, as well as detectability of exoplanets \citep{aigrain16}. 

Doppler Imaging (hereafter DI) is a widely used 
technique in reconstructing distributions of relative brightness (or temperature)
on the visible parts of rapidly rotating cool stars \citep[see,][]{Strassmeier2009}. 
For a realistic Doppler image, three critical prerequisites are (a) a sufficiently 
large rotation rate and a good 
coverage of rotational phases, (b) a precise determination of fundamental stellar 
parameters, (c) high-resolution spectra with sufficiently high signal-to-noise 
ratios. 

Using high-resolution spectra, detailed profiles of the chemical composition can 
also be obtained. This gives insight for studies of solar-like magnetic activity 
throughout stellar evolution, such as the Li abundance, which is often used 
as an indicator of the stellar age \citep[see, e.g.,][]{carlosetal16}. 
Elemental abundances are also important in revealing the initial chemical compositions of stellar birthplaces, and to evaluate the effects of metallicity on magnetic activity \citep{karoff18}. 

In this study, we infer the starspot distribution using time-series of high-resolution spectra. We also carry out an extensive chemical abundance analysis for a large number of elements aiming to derive the abundance pattern for the atmosphere of EK Dra, and compare this with the solar elemental abundances.
To interpret the observed Doppler images, we synthesise artificial ones, by applying the flux emergence and transport model developed by \citet{isik18}, considering the age of EK Draconis. 

\subsection{EK Dra}
\label{sec:ekintro}

EK Dra (G1.5V) can be considered as a laboratory to characterise the magnetic activity of a solar-type star at 
its early stages of evolution. Various ages between 30 and 125 Myr are proposed for EK Dra \citep[see, e.g.][]{soderblomandclements87,staufferetal98,wichmannetal03,Konig2005,Jarvinen2007}. It is the primary component of a 
widely separated binary system. The mass of the secondary component is about $0.5~M_\odot$ \citep{Konig2005}. EK Dra rotates at a rate of about 10 times that of the Sun (recently given by \citet{Jarvinen2018} as $P_{\rm rot}\sim 2.606$ km/s).

The first Doppler Image (hereafter DI) of 
this young solar analogue was performed by \citet{Strassmeier1998}, using high-resolution 
CFHT spectra obtained in 1995. They found several cool spots 
distributed throughout low- to mid-latitudes, while the dominant spot was 
located at around 70-80$^{\circ}$ latitudes. They estimated spots with 
temperature deficits in the range between 400-1200~K. Using photometric data 
gathered between 1994-1998, they pointed out that a 
high-latitude feature can be responsible for the continuous decline of the mean 
brightness. \citet{Frohlich2002} performed long-term photometry covering 35 
years, using Sky-Patrol plates obtained at Sonneberg Observatory and found a 
secular dimming of 0.0057 $\pm{0.0008}$ mag/yr within the observation time-range 
considered. \citet{Jarvinen2005} analyzed V-band photometric data of EK Dra covering 21 
years using light curve inversions and found that the system shows long-lived, 
non-axisymmetric spot distribution with active longitudes on opposite 
hemispheres. \citet{Konig2005} calculated the masses of the primary and 
secondary components as 0.9$M_{\rm\odot}$ and 0.5$M_{\rm\odot}$, 
respectively, and found that the orbit of the binary system is highly eccentric 
with $e=0.82$, using a combination of speckle observations and radial velocity 
measurements. They also obtained the effective temperature, $T_{\rm eff}$, as 5700 K and the surface gravity, $\log g$, as 4.37, using high-resolution spectra. In another 
detailed study of EK Dra, \citet{Jarvinen2007} found both high- and low-latitude spots 
500 K cooler than the quiet photosphere, using the inversions of atomic line profiles 
and mentioned both equatorward and poleward migration of spots around active longitudes. 
Furthermore, they pointed out that the equivalent widths of CaII-IRT emission cores  
increased during the light-curve minimum, indicating a positive correlation between photospheric 
and chromospheric activity. From a detailed model-atmosphere analysis of high-resolution spectra of EK Dra, they obtained $T_{\rm eff}$=5750 K, log\textit{g}=4.5, and the microturbulent velocity, $\xi$, as 1.6 km/s. They estimated an age between 30-50 Myr. 

\citet{Jarvinen2009} studied EK Dra, based on additional spectra obtained in 2007. They also inferred  high-latitude spot concentrations. \cite{Rosen2016} published two brightness maps, along with vector magnetic maps, using spectra obtained in 2007 and 2012. They found spots covering all longitudes and a wide latitudinal range, excluding high-latitude regions above $\sim 60^\circ$.

\citet{Waite2017} reconstructed brightness and magnetic maps using spectropolarimetric observations covering six years of data obtained at the CFHT and TBL observatories between 2006 and 2012. Their surface reconstructions between 2006 and 2007 show large spots located at intermediate latitudes, while the maps based on 2008 data show a dominant polar spot region. The maps also showed significant spot evolution over three-month periods, which the authors interpreted as rapid reorganisation of the global magnetic field. Using Stokes V data, they also obtained an average equatorial rotational velocity ($\Omega_{\rm eq}\sim 2.50 \pm 0.08~{\rm rad~d}^{-1}$) as well as a surface differential rotation rate with an average rotational shear of $\Delta\Omega \sim 0.27^{+0.24}_{-0.26}~{\rm rad~d}^{-1}$.

The most recent Doppler imaging study of EK Dra was carried out by \citet{Jarvinen2018} (hereafter J18), using a time series of very high resolution intensity spectra obtained at the Large Binocular Telescope equipped with the Potsdam Echelle Polarimetric and Spectroscopic Instrument (PEPSI). Using 10 spectra obtained between April 3 - 11 2015, they reconstructed a temperature map including 4 distinctive cool spots located at mid- to low latitudes, with temperature differences ranging from 280 K to 990 K with respect to the photosphere. 

There are only two studies conducting a detailed analysis of the abundances of several chemical elements for EK Dra. However, both of these studies are based on automated spectral synthesis modelling methods. \cite{valentietal05} derived the abundances of Na, Si, Ti, Fe, and Ni elements with respect to the solar abundances. [Na/H], [Si/H], and [Ni/H] are derived as -0.04 dex indicating solar-like values. However, they find [Ti/H]=$0.15\pm0.05$ and [Fe/H]=$0.16\pm0.03$, which are marginally larger than the solar abundances. \citet{Brewer2016} also derived the abundances of 15 elements. The abundances of C, Si, V, Cr, Mn, Fe, and Y are derived to be very close to the solar ones, i.e., within $\pm0.06$\,dex. However, they derived slightly larger abundances (0.11-0.22 dex) for N, O, and Ca, and slightly smaller abundances (between -0.13 and -0.19) for Na, Mg, Al, and Ni, with respect to the solar abundances. In addition to these studies, \citet{Konig2005}, \citet{Jarvinen2007}, and \citet{Jarvinen2018} reported ${\rm [Fe/H]} =-0.16\pm0.02$, $0.00\pm0.05$, and $-0.20\pm0.02$, respectively, from the spectrum synthesis. \citet{Konig2005} and \citet{Jarvinen2007} also derived the lithium abundance log\,$(N_{Li})$ for the star as $3.02\pm0.02$ and $3.30\pm0.05$, respectively.

\section{Observations and Data Reduction}
\label{sec:obs_and_red}

The high resolution time-series spectra of EK Dra were obtained between 17 and 31 July, 2015 with the CAFE spectrograph \citep{Aceituno2013} attached to the 2.2~m telescope at the Calar Alto Observatory (Almeria, Spain) and with the HERMES spectrograph \citep{Raskin2011} attached to the 1.2~m Mercator telescope at the Roque de los Muchachos Observatory (La Palma, Spain) between 15 December 2015 and 17 May 2016. During the Calar Alto run, we obtained 17 spectra covering a wavelength range between 4050$\mbox{\AA}$ - 9095 $\mbox{\AA}$ and an average resolution of R=62000, while we obtained 12 spectra of EK Dra with an average resolution of R=85000 and a wavelength coverage between 3780 - 9007 $\mbox{\AA}$. The log of observations including Signal-to-Noise Ratio (SNR) values is given in Table~\ref{tab:obslog}. We also observed slowly-rotating and non-active template stars HD~143761 (G0~V), HD146233 (G2~V), HD~4628 (K2.5~V) and HD~22049 (K2~V) during the observing runs. The ephemeris used for the phase calculation of the spectral data obtained in this study is taken from \citet{Jarvinen2018} and given in equation~\ref{ephe} below: 

\begin{equation}
{\rm HJD} = 2445781.8590 + 2^{\rm d}.606 \times E.
\label{ephe}
\end{equation}

As seen from Table~\ref{tab:obslog}, we have  good phase coverage of EK Dra; 17 spectra covering 15 days were  obtained from CAFE in July 2015. On the other hand, for the HERMES run, the spectral data covers 6 months distributed among December 2015, March 2016 and May 2016. Therefore, we used the CAFE data only for Doppler Imaging, while the HERMES data is used only for abundance analysis, since such intermittent coverage over such a length of time is not ideal for DI. In addition, the CAFE data used for the DI procedure almost uniformly samples the rotation period of the star, in fact it has the densest spectral coverage compared to previous DI studies. The reduction procedures such as bias and flat correction, removal of cosmic rays, wavelength calibration and heliocentric velocity correction for the Calar Alto data were performed with the help of the IRAF\footnote{IRAF is distributed by the National Optical Astronomy Observatory, which is operated by the Association of the Universities for Research in Astronomy, inc. (AURA) under cooperative agreement with the National Science Foundation.} (Image Reduction and Analysis Facility) standard packages. For the reduction process of HERMES data, we used the automatic pipeline of the spectrograph \citep{Raskin2011}.


We applied the signal enhancing Least Squares Deconvolution (LSD) technique \citep{Donati1997b} to obtain high SNR profiles of the spectral data. The LSD routine is based on a cross-correlation technique that combines photospheric lines in the observed spectrum into a single profile in the velocity space to enhance the SNR value. The line list required by the LSD technique was derived from the Vienna Atomic Line Database (VALD) \citep{kupkaetal99}, considering the effective temperature ($T_{\rm eff}$) and the surface gravity ($\log g$) of EK Dra. During the preparation of the line list, wavelength regions covering chromospheric lines as well as the strong telluric lines were extracted to prevent any artefacts in the LSD profiles. The LSD profiles of the template stars were also obtained, which are used for the generation of lookup tables to model the local intensity profile during DI process. The resolving power of R=62000 allows us to generate the velocity profiles with 2.5 km/s intervals.

\begin{table}
	\centering
	\caption{Phase-ordered spectroscopic observation log of EK Dra.}
	\label{tab:obslog}
	\begin{tabular}{lccr} 
		\hline
		Date & Exp. Time (sec.) & Phase$_{Mid}$ & SNR\\
		\hline
		\multicolumn{4}{|l|}{Calar Alto Observing run}\\
		\hline
        18.07.2015 & 1800 & 0.069 & 90 \\
        19.07.2015 & 1800 & 0.135 & 77 \\
        21.07.2015 & 2200 & 0.220 & 115 \\
        22.07.2015 & 2200 & 0.267 & 88 \\
        29.07.2015 & 1800 & 0.284 & 58 \\
        30.07.2015 & 1800 & 0.339 & 69 \\
        19.07.2015 & 1800 & 0.495 & 90 \\
        27.07.2015 & 1800 & 0.536 & 83 \\
        22.07.2015 & 2300 & 0.597 & 115 \\
        22.07.2015 & 2300 & 0.644 & 80 \\
        30.07.2015 & 1800 & 0.671 & 90 \\
        17.07.2015 & 1800 & 0.679 & 81 \\
        31.07.2015 & 1800 & 0.720 & 70 \\
        20.07.2015 & 2000 & 0.834 & 103 \\
        21.07.2015 & 2200 & 0.885 & 93 \\
        28.07.2015 & 1800 & 0.909 & 93 \\
        29.07.2015 & 1800 & 0.961 & 55 \\

		\hline
		\multicolumn{4}{|l|}{La Palma Observing run}\\
		\hline
        25.03.2016 & 1200 & 0.107 & 131 \\
        17.12.2015 & 1200 & 0.164 & 134 \\
        28.03.2016 & 1200 & 0.259 & 154 \\
        16.05.2016 & 1200 & 0.359 & 100 \\
        15.12.2015 & 1200 & 0.401 & 96 \\
        26.03.2016 & 1200 & 0.491 & 109 \\
        29.03.2016 & 1200 & 0.646 & 155 \\
        17.05.2016 & 1200 & 0.735 & 147 \\
        27.03.2016 & 1000 & 0.870 & 145 \\
        19.12.2015 & 1200 & 0.939 & 111 \\
        15.05.2016 & 1800 & 0.963 & 96 \\
        15.05.2016 & 1200 & 0.985 & 107 \\
  
		\hline
	\end{tabular}
\end{table}

\section{Fundamental Parameters}
\label{sec:fundapars}

We have estimated the effective temperature ($T_{\rm eff}$), surface gravity ($\log g$), rotational velocity ($v{\rm sin}i$), microturbulence ($\xi$), distance, mass ($M$), radius ($R$), and age ($\tau$) of EK Dra using various photometric and spectroscopic methods. The results and adopted values are summarised in Table \ref{tab:atmpar}. The uncertainties of the adopted atmospheric parameters were estimated from the standard deviations of the individual measurements. 

\begin{table}
	\centering
	\caption{Fundamental Parameters of EK Dra.}
	\label{tab:atmpar}
	\begin{tabular}{lll} 
		\hline
        \hline
		Parameter & Value$\rm{^a}$ & Method and notes \\
		\hline
		$T_{{\rm eff}}\ [K]$  & $5672$ & Str\"{o}mgren photometry \\
		                 & $5811\pm65$ & Geneva photometry \\
    	                 & $5799\pm100$ & Johnson photometry + F96 \\
		                 & $5926\pm100$ & Johnson photometry + B98 \\
		                 & $5776\pm100$ & Johnson photometry + SF00 \\
		                 & $5770\pm120$ & {\it SDSS} $g$ and $r$ filters + B12 \\
		                 & $5664\pm120$ & {\it Gaia} photometry + B12 \\
		                 & $5770\pm50$ & Excitation eq. of Fe {\small I} \\
		                 & $ 5770\pm80$ & Adopted \\
		Log\,$g$ [cgs]   & $4.41$ & Str\"{o}mgren photometry \\
		              & $4.28\pm0.26$ & Geneva photometry \\
    	              & $4.54\pm0.15$ & {\it SDSS} $g$ and $r$ filters + B12 \\
		              & $4.46\pm0.12$ & {\it Gaia} photometry + B12 \\
		              & $4.40\pm0.10$ & Ion. eq. of Ti I/II and Cr I/II  \\
		              & $4.4\pm0.1$ & Adopted \\
        $\xi_{{\rm mic}}$ [km\,s$^{-1}$] & $1.7\pm0.1$ & by minimizing $\sigma_{{\rm [Fe/H]}}$ \\
		Dist. [pc]  & $34.45\pm0.03$ & {\it Gaia} parallax \\
		$M$ [$M_{\rm \odot}$]  & $1.04\pm0.04$ & $T_{\rm eff} - {\rm log}g$ diagram + B12 \\
		$R$ [$R_{\rm \odot}$] & $1.07\pm0.10$ & $T_{\rm eff} - {\rm log}g$ diagram + B12 \\
		Log\,$\tau\rm{^b}$  & $\lesssim7.7$ &  From Li abundance \\
         & $7.43\pm0.15$ & $T_{\rm eff} - {\rm log}g$ diagram + B12 \\	
    		              &  & (Adopted) \\
		vsin$i$ [km\,s$^{-1}$]  & $17.0\pm0.5$ & From Fe I lines \\
		                        & $16.6\pm0.2$ & From DI (Adopted) \\
		v$_{{\rm rad}}$ [km\,s$^{-1}$]  & $-20.26\pm0.01$  & From DI \\
		$i$ [$^\circ$]  & $63$  & From J18 \\
		\hline
        \hline
	\end{tabular}
	    \\
	    \begin{flushleft}
	    \footnotesize{$\rm{^a}$ Systematic uncertainties caused by isochrones (B12) were assumed as $\pm 0.1$ dex for log\,$g$ and $\pm100$ K for $T_{{\rm eff}}$.} \\
        \footnotesize{$\rm{^b}$ $\tau$: age in years.} 
        \end{flushleft}
\end{table}

The magnitudes of EK Dra in Johnson ($BV$), Geneva ($UB_1B_2V_1G$), and Str\"{o}mgren ($uvby$) systems were collected from Hipparcos/Tycho catalogue \citep{ESA97}, \citet{rufener88}, \citet{olsen83}, and \citet{hauckandmermilliod98}. $T_{\rm eff}$ and log\,$g$ of the star were initially derived using the photometric calibrations of  \citet[F96]{flower96}, \citet[B98]{besselletal98}, \citet[SF00]{sekiguchiandfukugita00}, \citet[K97]{kunzlietal97},  and \citet[N97]{napiwotzki97}. The distance of the star was calculated to be $35\pm0.4$ pc from its {\it Gaia} DR2 parallax \citet{Gaia2016,Gaia2018}. The colour excess of the star was derived to be $E(b-y)=0.002$ from Str\"omgren photometry. At the distance and galactic position of EK Dra, the 3D dust map of \citet{green15} also gave a null (i.e., $E(B-V)=0.00$) colour excess. The colour excess of the star was therefore neglected for all photometric systems.

\begin{figure}
	\includegraphics[width=\columnwidth]{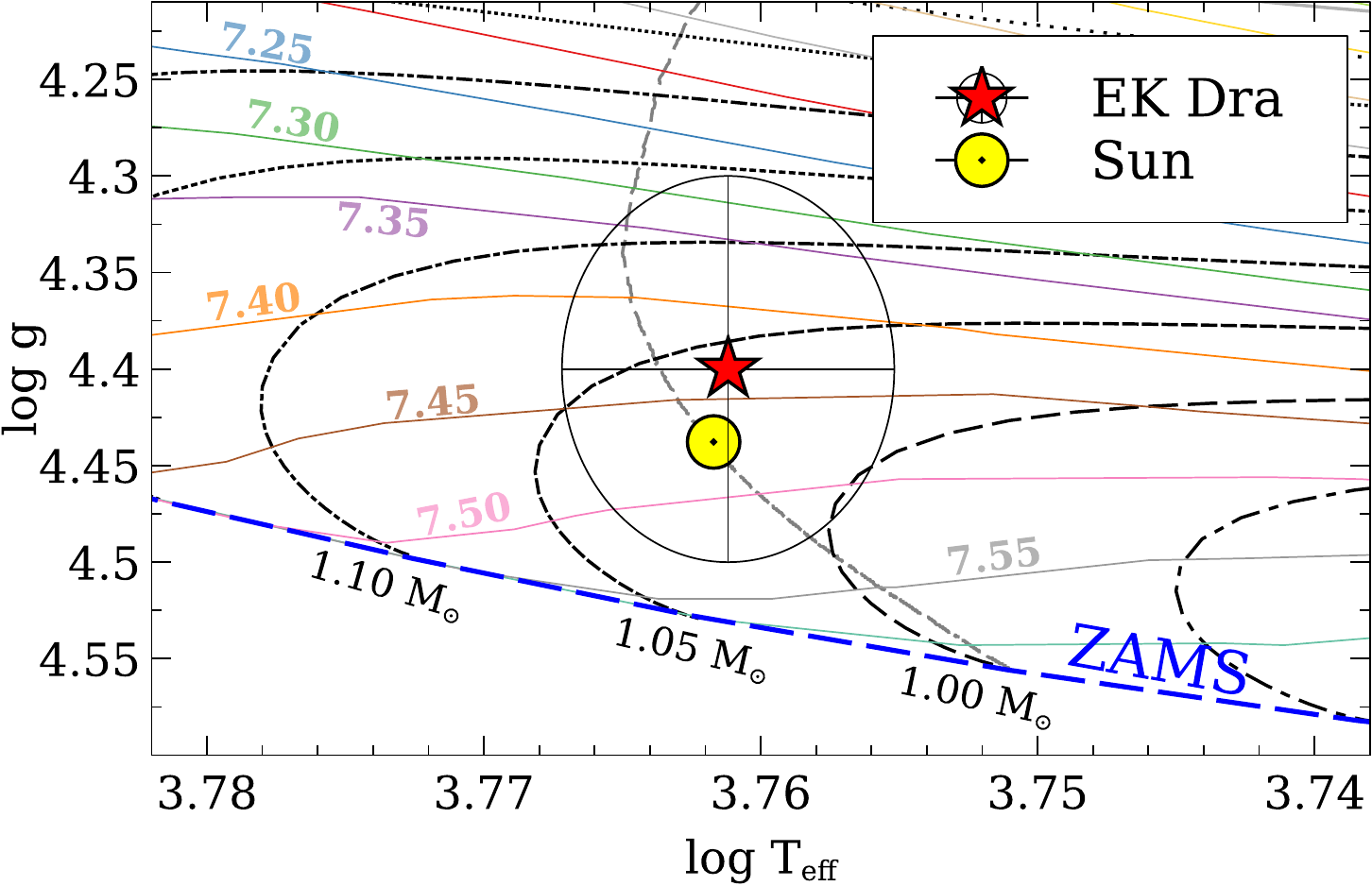}
    \caption{EK Dra on a $T_{\rm eff}-{\rm log}g$ diagram with {\scriptsize PARSEC} isochrones and pre-main sequence (PMS) evolutionary tracks. The masses are shown below each evolutionary track. The logarithm of the ages (log\,$\tau$, $\tau$ in years) are also shown above each isochrone with corresponding colour codes (we note that these ages are only valid for PMS stars). The theoretical position of the Sun and its main-sequence evolutionary track (dashed gray line) are displayed only for a comparison purpose. All tracks and isochrones are for the initial metallicity $z=0.01774$.}
    \label{fig:evol}
\end{figure}

Besides the photometric calibrations in Str\"omgren and Geneva systems, we also estimate the atmospheric parameters of EK Dra using colour-magnitude diagrams with theoretical isochrones for {\it Gaia} DR2 (G, $G_{\rm BP}$ and $G_{\rm RP}$) and SDSS DR14 ($gr$) magnitudes. The isochrones were taken from PARSEC \citep[B12]{bressanetal12}. We chose the isochrones computed for \citeauthor{weiler2018}'s (\citeyear{weiler2018}) {\it Gaia} filter transmissions proposed for bright stars. The mean of these photometrically estimated $T_{\rm eff}$ and $\log g$ values were refined further through spectroscopic methods (see Sec. \ref{sec:chemcomp} for the details) and the final atmospheric parameters were derived as $T_{{\rm eff}}=5770\pm80$\,K and log\,$g=4.4\pm0.1$.  We then plotted the star on the $T_{\rm eff}-{\rm log}g$ diagram (Fig. \ref{fig:evol}) and estimated the mass, radius, and the age of the star. A mass of $1.04\pm0.04$ M$\odot$ and an age of log\,$\tau = 7.43 \pm 0.15$ (i.e., $27^{+11}_{-8}$\,Myr) were consequently estimated, putting EK Dra on the pre-main sequence phase of its evolution, most likely in the post-T Tauri stage. 
 
\section{Chemical Abundance Analysis}
\label{sec:chemcomp}

The model atmospheres were computed with ATLAS12 \citep{kurucz93, sbordoneetal04} assuming Local Thermodynamic Equilibrium (LTE) and plane-parallel geometry. For the spectrum synthesis, the atomic data was retrieved from VALD \citep{piskunovetal95, ryabchikovaetal97, kupkaetal99, kupkaetal00} and the NIST database \citep{NIST_ASD}. The detailed list of the atomic lines can be found in \citet{kilicogluetal16}. The oscillator strengths of $^7$Li lines were adopted from \citet{Reddy2002}. The mixing length parameter were calculated to be $\alpha=1.53$ from the calibration of \citet{Ludwig1999}.

In order to avoid photospheric activity as much as possible, we selected spectra obtained by the HERMES spectrograph when the star is in a less active phase, i.e., the spectra have weak emission features in the Ca II H\&K lines. Three out of 12 observed HERMES spectra, obtained consecutively on March 29th, 2016, fitted this condition. We merged these three spectra to increase the S/N and used the merged spectrum for chemical abundance analysis. We carefully selected 434 unblended (or barely blended) lines of Li, C, Na, Mg, Al, Si, S, Ca, Sc, Ti, V, Cr, Mn, Fe, Co, Ni, Cu, Zn, Sr, Y, Ba, Ce, and Nd in the spectrum. We then  fitted these observed spectra line-by-line with synthetic profiles produced by SYNSPEC49 and its SYNPLOT interface. The code was modified to minimise the $\chi^{2}$ between the model and the observed points, using the Levenberg-Marquardt algorithm \citep{markwardt09}. The rotational velocity was derived to be $v$sin$i$ = $17.0\pm0.5$~km\,s$^{-1}$ by adjusting the synthetic spectra to the unblended iron lines.

\begin{figure}
	\includegraphics[width=\columnwidth]{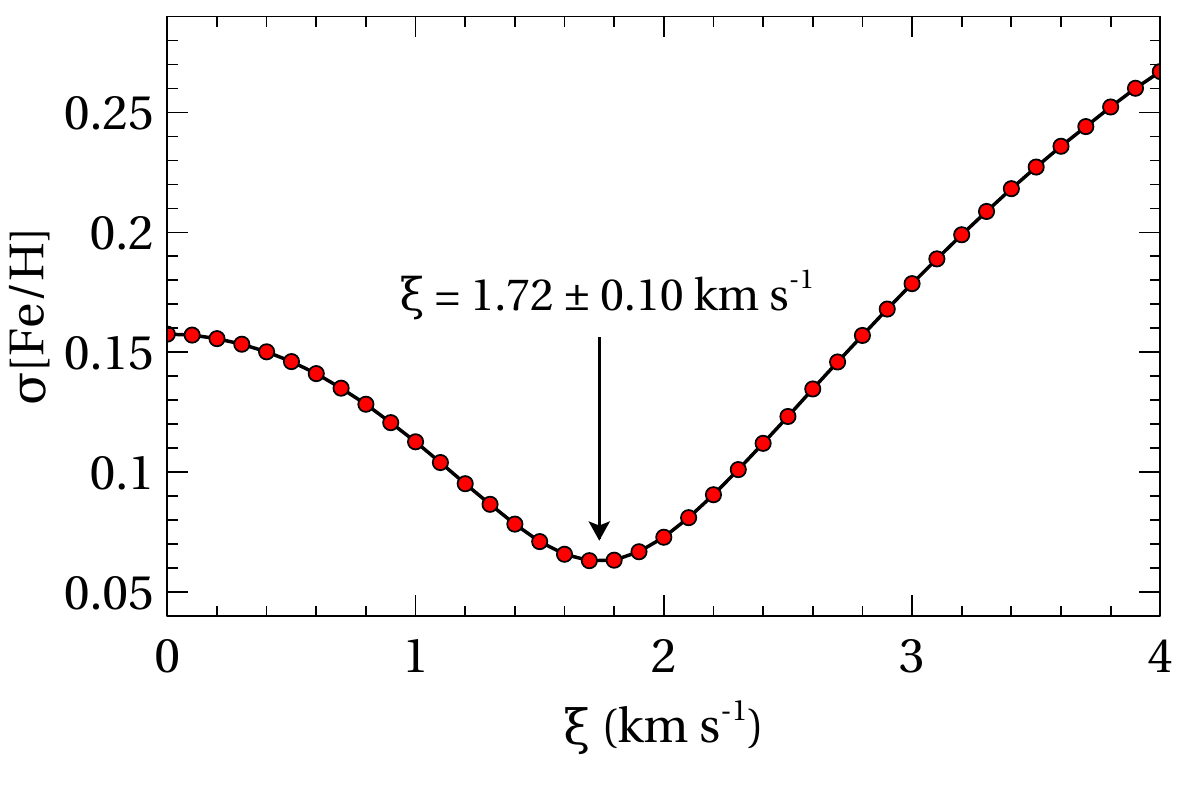}
    \caption{Standard deviation of the derived iron abundance vs. the microturbulent velocity for EK Dra.}
    \label{fig:microturb}
\end{figure}

To derive the microturbulent velocity of EK Dra, we obtained the iron abundance [Fe/H] using 190 unblended Fe~I lines for a set of microturbulent velocities ranging from 0.0 to 4.0 km\,s$^{-1}$. Figure~\ref{fig:microturb} shows the standard deviation of the derived ${\rm [Fe/H]}$ as a function of the microturbulent velocity. The adopted microturbulent velocity is the value which minimises the standard deviation, i.e., $\xi = 1.72 \pm 0.10$ km\,s$^{-1}$. The abundances of the other elements were determined after this step using the obtained $\xi$ value.   

\begin{figure}
	\includegraphics[width=\columnwidth]{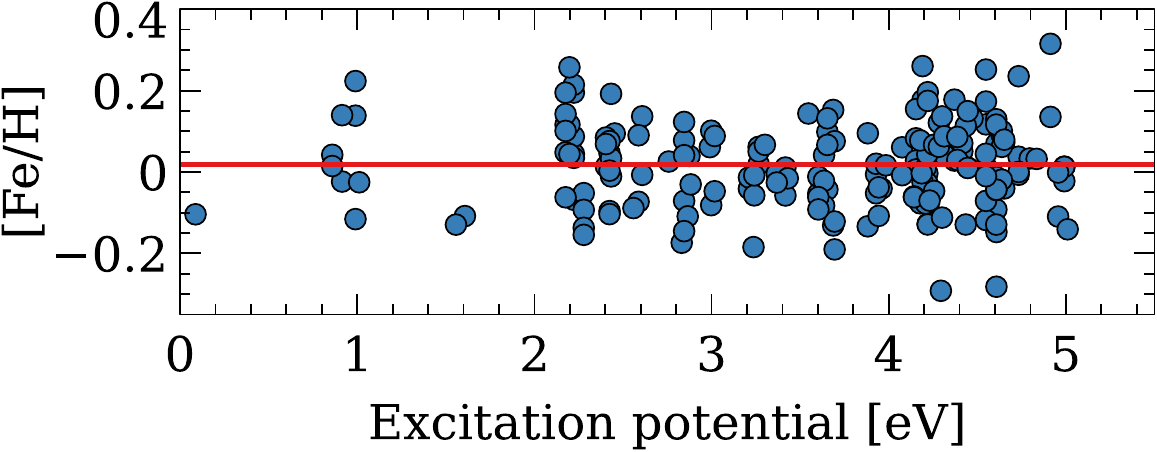}
    \caption{[Fe/H] versus excitation energies of the Fe I lines for $T_{\rm eff}=5770$\,K. Red regression line with a slope of 0.0 confirms the excitation equilibrium.}
    \label{fig:excitation}
\end{figure}

The atmospheric parameters of EK Dra were refined by testing the excitation equilibrium of Fe I and the ionisation equilibrium of Ti I/II and Cr I/II. The scatter plot of excitation potential vs. Fe abundances (from Fe I lines) gave the minimum regression slope at $T_{\rm eff} = 5770$\,K (Fig. \ref{fig:excitation}). For log\,$g$=4.4, we derived the same Cr abundance from both Cr I and Cr II lines, and the same Ti abundances from both Ti I and Ti II lines (within $\pm0.02$\,dex). We could not use Fe lines for ionisation equilibrium because there were an insufficient number of Fe II lines in the spectra for the analysis. These atmospheric parameters therefore provided reasonable ionisation and excitation equilibria for the star.

We performed a detailed uncertainty analysis for the chemical abundances. Within the uncertainty limits, we altered $T_{\rm eff}$, $\log g$, and $v_{\rm mic}$ parameters of the synthetic spectrum and re-derived the abundances for each variation. We also derived standard errors ($\sigma^2_{\rm sterr}$) of the abundances derived from individual lines for each element. Therefore, the uncertainties given in Table \ref{tab:abun} are calculated using the assumption below:


\[
\sigma_{\rm tot} = \sqrt{\sigma^2_{T_{\rm  eff}}+\sigma^2_{{\rm log}g}+\sigma^2_{\rm \xi}+\sigma^2_{\rm sterr}}
\]

\begin{figure*}
	\includegraphics[width=\textwidth]{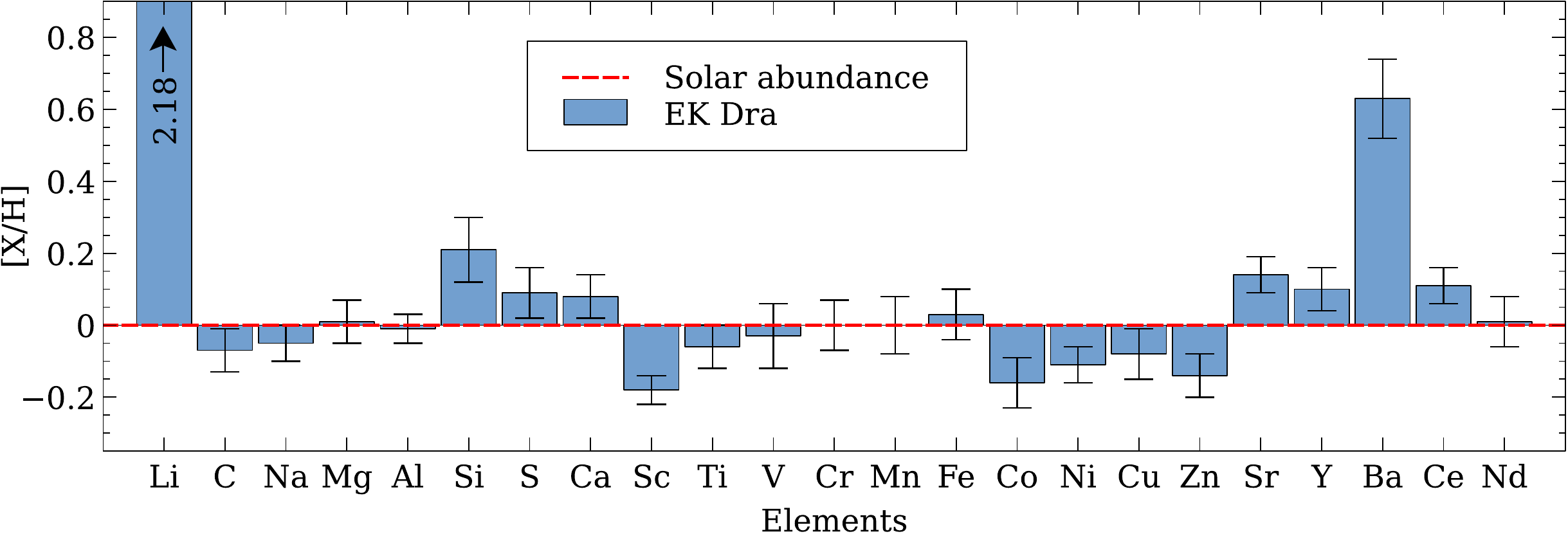}
    \caption{Logarithmic abundances of the elements for EK Dra relative to their solar abundances.}
    \label{fig:abundances}
\end{figure*}

\begin{figure}
	\includegraphics[width=\columnwidth]{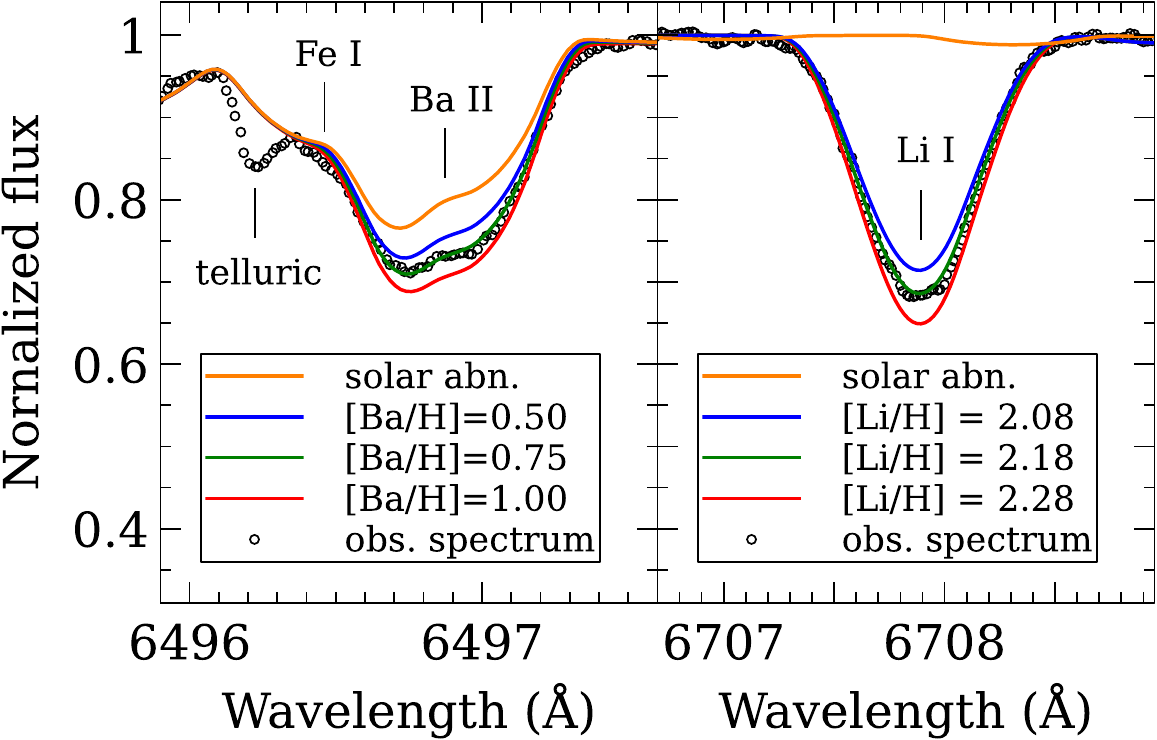}
    \caption{Ba\,{\scriptsize II} 6496.989 and L\,{\scriptsize I} 6707.983 lines yielding overabundances.}
    \label{fig:liba}
\end{figure}

The final abundances of 23 elements are given in Table \ref{tab:abun} and the deviations from the solar abundances \citep[taken from ][]{grevessesauval98} are illustrated in Fig. \ref{fig:abundances}. We have detected significant overabundances of Li and Ba (Fig. \ref{fig:liba}). The 3D non-LTE (NLTE) correction for Li abundance $\Delta A({\rm Li})=0.07$ has been adopted from \citet{harutyunyan18}\footnote{We note that their NLTE corrections are valid for $5800\,{\rm K}\,\leq\,T_{\rm eff}\,\leq\,6500\,{\rm K}$ and $1.0 \leq A({\rm Li}) \leq 2.7$ and we slightly extrapolate it for $T_{\rm eff} = 5770\,{\rm K}$ and $A({\rm Li})=3.28$}. The lithium abundance $A_{\rm NLTE}({\rm Li})$\footnote{$A({\rm X}) = {\rm log}(N_{\rm X}/N_{\rm H})+12$}$=3.35$ we measure is quite large and comparable with that of T Tauri stars. This shows that there is no remarkable Li depletion for EK Dra. \citet{Martin92} have reported that most of the lithium depletion occurs in the last stage of pre-main-sequence evolution, i.e., about 50 Myr for $1.0\,M_\odot \lesssim M \lesssim 1.1\,M_\odot$. Therefore, the photospheric lithium abundance of EK Dra indicates that the star is most likely younger than $\sim{}50$\,Myr, and this age limit agrees well with the age we found from the $T_{\rm eff}-{\rm log}g$ diagram ($27^{+11}_{-8}$\,Myr). The cause of the Ba overabundances seen in very young chromospherically active stars has recently been investigated by \citet{reddyandlambert17}. They concluded the Ba abundances could be overestimated due to the depth-independent microturbulence approximation: the line forming layer of Ba II is much higher than that of Fe, and consequently, larger microturbulent velocities are needed to model the Ba II lines. In their Fig 10., stars with larger chromospheric activity index log\,$R'_{\rm HK}$ tend to exhibit larger Ba abundances (from Ba II lines). The activity index of EK Dra has been reported to be log\,$R'_{\rm HK} = -4.19 \pm 0.04$\footnote{averaged log\,$R'_{\rm HK}$ of the published maximum and minimum values with deviation} by \citet{Rosen2016} and log\,$R'_{\rm HK} = -4.08 \pm 0.05$\footnote{averaged log\,$R'_{\rm HK}$ of the published three values with standard deviation} by \citet{borosaikiaetal2018}. Using a rough extrapolation from the Fig 10. of \citet{reddyandlambert17} for the mean log\,$R'_{\rm HK} \approx -4.13$, we reach [Ba/H] $\approx +0.45$ which is indeed slightly below the Ba abundance we found from Ba II lines\footnote{The resonance lines of Ba II (at 4554 and 4934 \AA) not included in the analysis.}, i.e., [Ba/H]\,$=0.63$. This makes the photospheric Ba abundance of EK Dra only slightly larger than solar (i.e., [Ba/H]\,$\approx 0.18$) and quite similar to the other adjacent heavy element abundances, such as, Sr, Y, and Ce. The abundances of all studied elements heavier than Li deviate by $\pm 0.21$ or less from the solar values. The largest deviation (except for Li and Ba) belongs to Si. However, the Si abundance was derived only from two lines and it might be affected by the random errors of the oscillator strengths. The abundances of Co, Ni, Cu, and Zn is slightly lower and the abundances of Sr, Y, Ba, and Ce are slightly greater than the solar photospheric abundances. The reason of these deviations are discussed in Sec.\,\ref{sec:discuss}.

\begin{table}
	\centering
	\caption{Abundances of the elements for EK Dra.}
	\label{tab:abun}
	\begin{tabular}{lrrlrr} 
		\hline
        \hline
		Element & log$(N_{\rm X}/N_{\rm H})$ & [X/H]* & \# of lines  \\
		        & $+12$                            &       &              \\
		\hline
		${\rm Li}_{\rm LTE}$ & $3.28\pm0.08$ & $2.18$ & 1 (Li I) \\
		${\rm Li}_{\rm NLTE+3D}$ & $3.35\pm0.08$ & $2.25$ & 1 (Li I) \\
		C &  $8.45\pm0.06$ & $-0.07$ & 4 (C I) \\
		Na & $6.28\pm0.05$ & $-0.05$ & 3 (Na I) \\
        Mg & $7.59\pm0.06$ & $0.01$ & 3 (Mg I) \\
        Al & $6.46\pm0.04$ & $-0.01$ & 4 (Al I) \\
        Si & $7.76\pm0.09$ & $0.21$ & 2 (Si II) \\
        S & $7.42\pm0.07$ & $0.09$ & 5 (S I) \\
        Ca & $6.44\pm0.06$ & $0.08$ & 18 (Ca I/II) \\
        Sc & $2.99\pm0.04$ & $-0.18$ & 4 (Sc I/II) \\
        Ti & $4.96\pm0.06$ & $-0.06$ & 41 (Ti I/II) \\
        V & $3.97\pm0.09$ & $-0.03$ & 16 (V I) \\
        Cr & $5.67\pm0.07$ & $0.00$ & 51 (Cr I/II) \\
        Mn & $5.39\pm0.08$ & $0.00$ & 7 (Mn I) \\
        Fe &  $7.53\pm0.07$ & $0.03$ & 190 (Fe I) \\
        Co & $4.76\pm0.07$ & $-0.16$ & 8 (Co I) \\
        Ni & $6.14\pm0.05$ & $-0.11$ & 58 (Ni I) \\
        Cu & $4.13\pm0.07$ & $-0.08$ & 2 (Cu I) \\
        Zn & $4.46\pm0.06$ & $-0.14$ & 2 (Zn I) \\
        Sr & $3.11\pm0.05$ & $0.14$ & 3 (Sr I/II)\\
        Y & $2.33\pm0.06$ & $0.10$ & 3 (Y II) \\
        Ba & $2.76\pm0.11$ & $0.63$ & 4 (Ba II) \\
        Ce & $1.69\pm0.05$ & $0.11$ & 2 (Ce II) \\
        Nd & $1.51\pm0.07$ & $0.01$ & 3 (Nd II)\\
        
		\hline
        \hline
        \multicolumn{6}{l}{*[X/H]$={\rm log}(N_{\rm X}/N_{\rm H}) - {\rm log}(N_{\rm X,\odot}/N_{\rm H,\odot})$\ \ \ \ \ \ \ \ \ \ \ \ \ \ \ \ \ \ \ \ \ \ \ } \\

	\end{tabular}
\end{table}

\begin{figure}
	\includegraphics[width=\columnwidth]{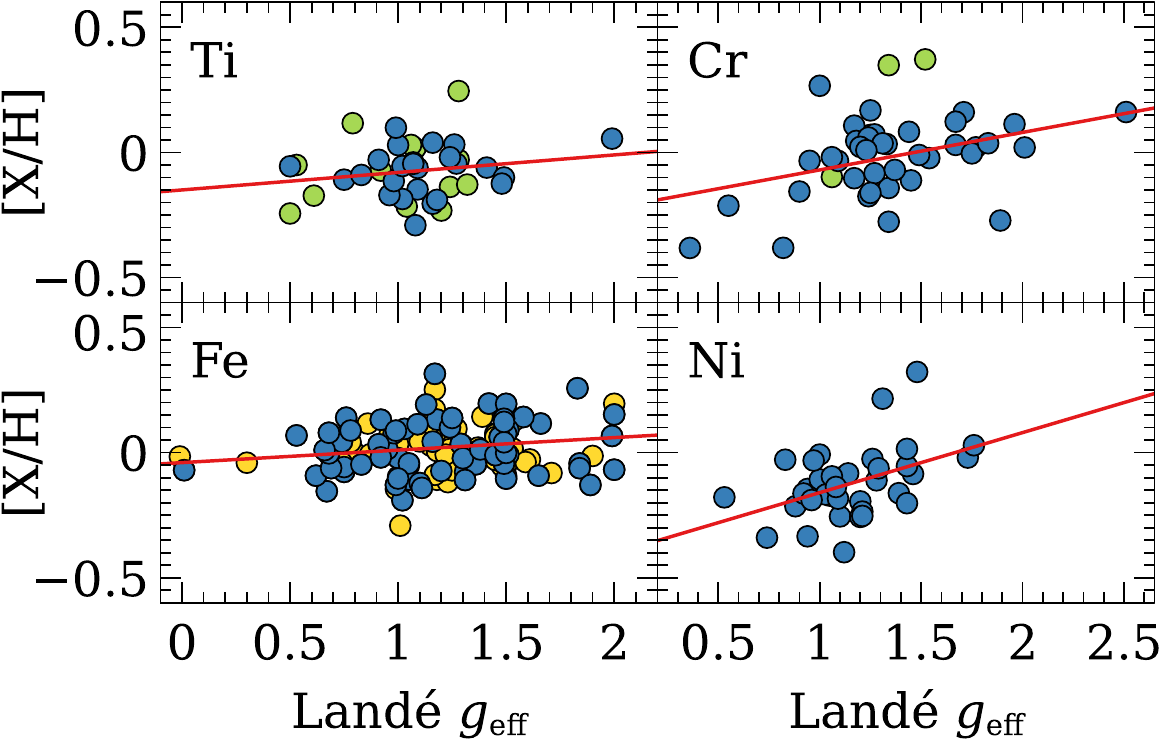}
    \caption{Line-by-line abundances of the elements versus effective Land\'e factors ($g_{\rm eff}$) of the lines. The abundances derived from the first and the second ionisation stages of the elements are represented by blue and green circles, respectively. For Fe, the yellow circles also represent Fe I, but they are for the lines with lower quality of oscillator strengths, i.e., quality of B+, B or worse in the accuracy classification of NIST.}
    \label{fig:landeg}
\end{figure}

The lowest and highest values for the magnetic field $\langle B \rangle$ of EK Dra were reported as 57 and 92 G, respectively, by \citet{Waite2017}. To test whether this magnetic field affects the derived chemical abundances, we plotted the abundances derived from individual lines with their effective Land\'e $g$ factors for Ti, Cr, Fe, and Ni (Fig. \ref{fig:landeg}). Here, we note that we do not consider the change in the thermal properties of the stellar atmosphere introduce by magnetic fields \citep[e.g.,][]{solanki86,solankiandbrigljevic92,solanki93}, which can influence the strength of spectral lines irrespectively of their Land\'e $g$ factors \citep[e.g.,][]{solankiandstenflo84,shelyagetal07}. We also note that the polarimetric estimates of $\langle B \rangle$ probably underestimate the actual values, because small-scale magnetic flux elements remain undetected. \citet{seeetal19}, however, suggested that the level of underestimation may be minor for the specific case of EK Dra. Fig. \ref{fig:landeg} indicates the lines with larger Land\'e $g$ factors tend to give larger abundances, most-likely due to the magnetic strengthening of the absorption lines. This effect appears to be negligible for Ti and Fe with a Pearson Correlation Coefficient (PCC) of about 0.14 and the slopes of their regression lines are only 0.07 and 0.05, respectively. However, the effect is more noticeable for Cr, and particularly for Ni, with a larger PCCs of 0.29 and 0.45, and larger slopes of 0.15 and 0.24, respectively. This shows that our derived Cr snd Ni abundances might be slightly overestimated due to the magnetic field. Our test is restricted to only four chemical elements as the other elements do not provide sufficient number of modelable lines. 


\section{The starspot distribution}
\label{sec:paramsdi}

\subsection{Doppler Imaging}

It is known that inaccurate stellar parameters lead to artefacts in the reconstructed Doppler images \citep[e.g.][]{Unruh1996}. Optimal stellar parameters are thus crucial for DI reconstructions. Here, we take advantage of the fact that several DI reconstructions of EK Dra are available in the literature, providing us with a reliable set of initial parameters to use in our analysis. In this section, we first explain how we adopt the parameters obtained in this study (see Sections \ref{sec:fundapars} and \ref{sec:chemcomp}), and carry out a 2D grid search for the critical parameters used in the DI procedure.

We used the Doppler imaging code {\tt DoTS} \citep{Cameron1997,Cameron1992}, which is based on the Maximum Entropy Method (MEM) for iterative regularisation of the resulting image. The {\tt DoTS} code reconstructs the spot filling factor ($f_{s}$) across the observable part of the star, using a two-temperature approximation for the representation of the quiet photosphere and spotted pixels. This procedure requires look-up tables generated using spectra of template stars for the calculation of intensity contributions of each element across the stellar surface. We assumed the temperatures of the quiet and spotted photospheres as 5750 K and 5000 K, respectively, by adopting the average spot temperature from J18. We used the linear limb-darkening coefficients derived by \citet{Claret2012, Claret2013}. 

\begin{figure}
	\includegraphics[width=\columnwidth]{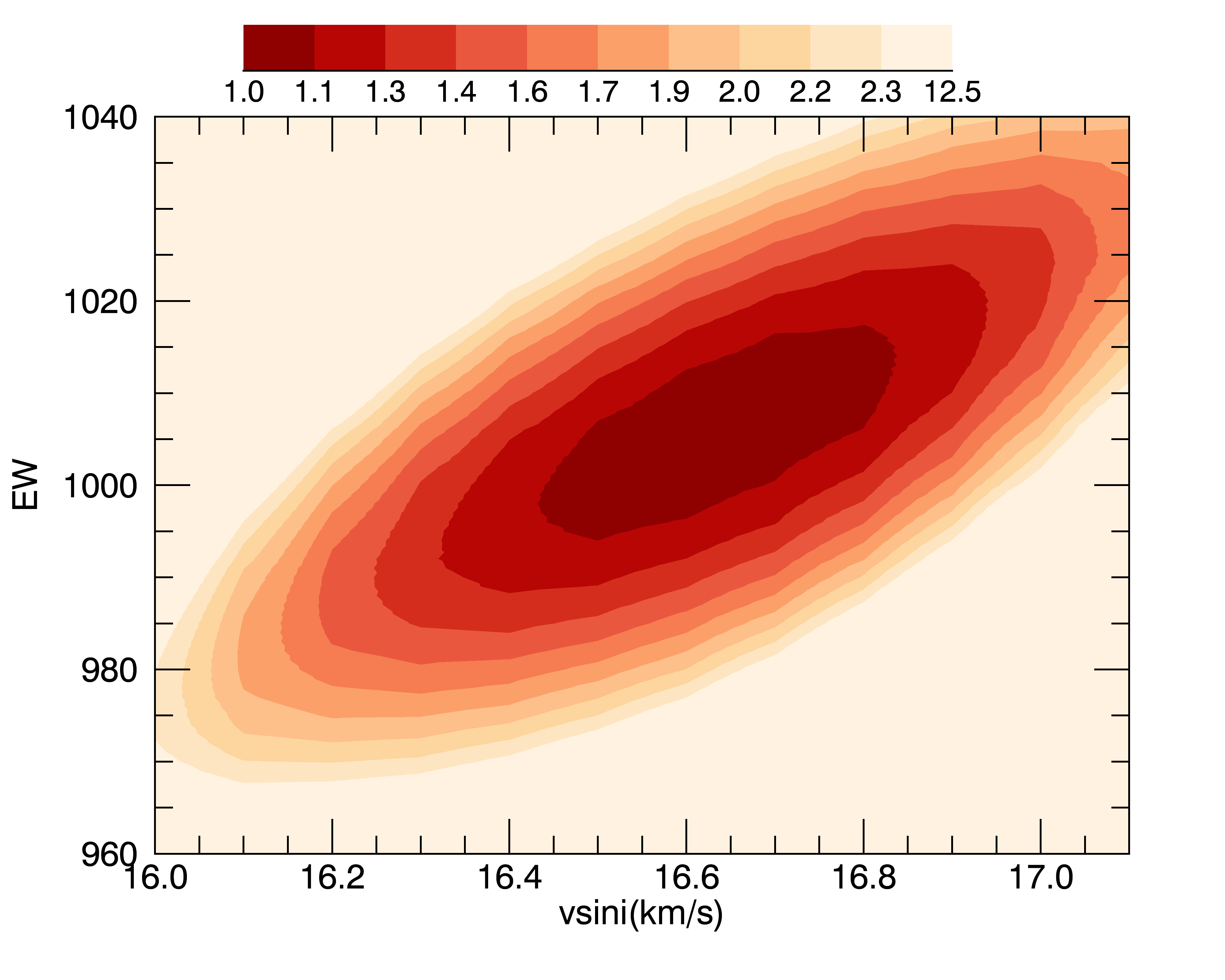}
   \caption{Two-dimensional grid search showing reduced $\chi^2$, used to optimise \textit{EW} and ${v \sin i}$ parameters.
   }
    \label{fig:gridsearch}
\end{figure}

For the input stellar parameters required by the {\tt DoTS} code, we used the ones obtained in Section~\ref{sec:fundapars}, except for the axial inclination $i$ as $63^\circ$, the rotation period and the epoch as given in Equation~\ref{ephe}, which we took from J18. In order to avoid spurious reconstruction of high-latitude spots, we performed a two-dimensional grid search for the equivalent-width parameter \textit{EW} vs. $v\sin i$. The \textit{EW} parameter controls the strength of the synthetic line profile. The resulting $\chi^2$ values are shown in Fig.~\ref{fig:gridsearch}. The $v\sin i$ value of 16.6$\pm0.2$ km/s obtained from the grid search is in accordance with both J18 and the one based on the spectral synthesis in Sect.~\ref{sec:fundapars} (see Table~\ref{tab:atmpar}). The error of the $v\sin i$ parameter was estimated from $\chi^2_{\rm min} + 1$ of $v\sin i$ value obtained from two-dimensional grid search \citep[see][for details]{Bevington2003}. Following MEM iterations, we achieved a reduced $\chi^2$ value of 1.127. In addition to the two-dimensional grid search for the \textit{EW} and $v\sin i$ parameters (Fig.~\ref{fig:gridsearch}), we also carried out a grid search for the axial inclination (see Appendix \ref{sec:appena} for details). Although the search result gives an axial inclination value of $56^\circ \pm 5^\circ$, we keep the value as $63^\circ$  during DI process for a more consistent comparison of the resultant map with the one obtained by J18 (see Section \ref{ssec:compare}). The  reconstructed spot positions and sizes do not alter significantly between inclination angles of $56$ and $63^\circ$.

\begin{figure}
	\includegraphics[width=\columnwidth]{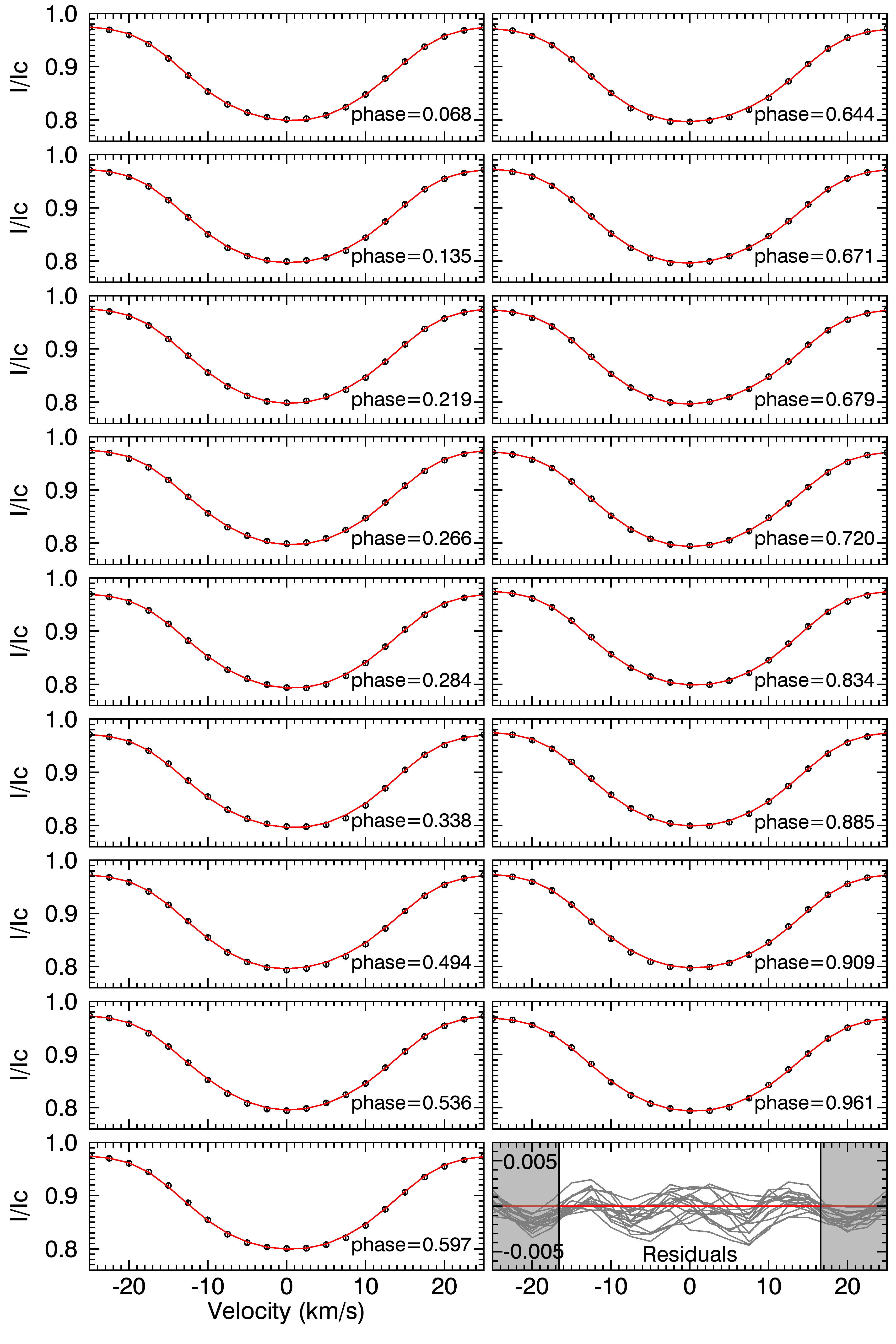}
    \caption{Phase-ordered LSD profiles (black circles). Red solid lines represent the maximum-entropy regularised models with the wavelength expressed in terms of radial velocity. The lowest-right panel shows the residuals of all observed profiles from the synthesised profiles, where the shaded regions mark the velocities beyond the $\pm v\sin i$ value. The rms value obtained from the combined residuals is $1.57\times 10^{-3}$ and $\chi^2= 5.0\times 10^{-5}$. Note that the error bars of each data point remain inside the black circles.}
    \label{fig:lsdfits}
\end{figure}

The best-fit models for the velocity profiles obtained using the Calar-Alto dataset are given in Fig.~\ref{fig:lsdfits}, along with the residuals, which are within $\pm 0.0035$. The residuals are likely related to uncertainties in the estimation of stellar parameters. In order to examine the ${v \sin i}$ value that is obtained from $\chi^2$ minimisation, we also performed an alternative approach previously carried out by \citet{Donati2003} and recently by \citet{Cang2020}, which is based on optimising the ${v \sin i}$ (and the \textit{EW} parameter in our case) until phase-averaged residuals are minimised (see Appendix \ref{sec:appenb} for details). The resulting surface reconstructions (with the obtained $v\sin i$ of 16.6 km/s) are indistinguishable from our map in Fig.~\ref{fig:lsdfits}. 

\begin{figure*}
	\includegraphics[width=40pc,height=10pc]{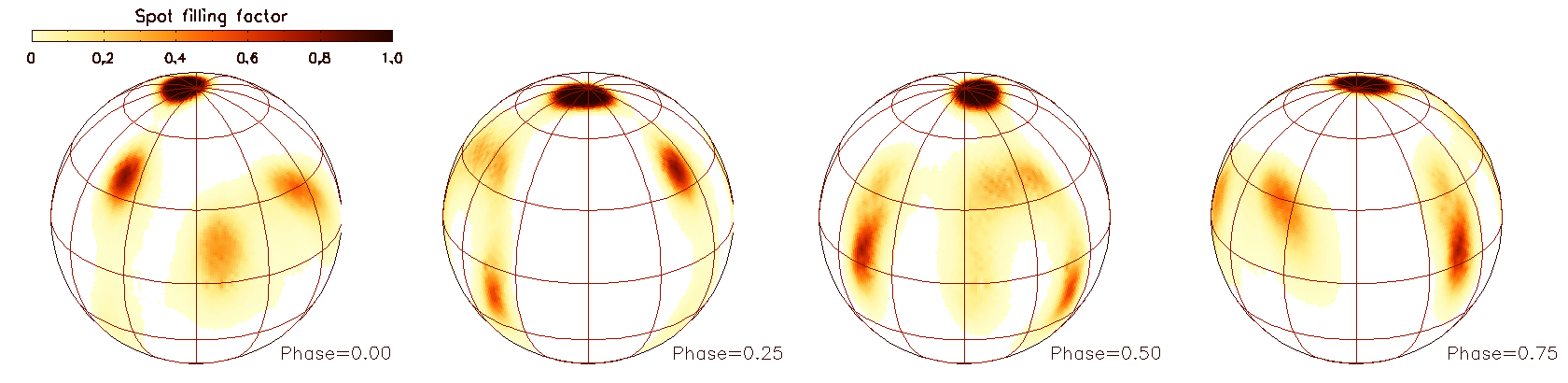}
    \caption{Projected disc images of EK Dra obtained from DI inversions at different rotational phases $\phi$=0.0, 0.25, 0.50, and 0.75.}
    \label{fig:sphmap}
\end{figure*}

The projected disc images corresponding to the resulting Doppler image are shown in Fig. \ref{fig:sphmap} at different rotational phases. The highest mean filling factor pertains to the spot near the pole, centred at around $77^\circ$ latitude. The remaining spots are distributed from middle to low latitudes. A detailed discussion on the spot distribution will be given in Sections \ref{ssec:compare} and \ref{ssec:spots}.

\begin{figure}
	\includegraphics[width=\columnwidth]{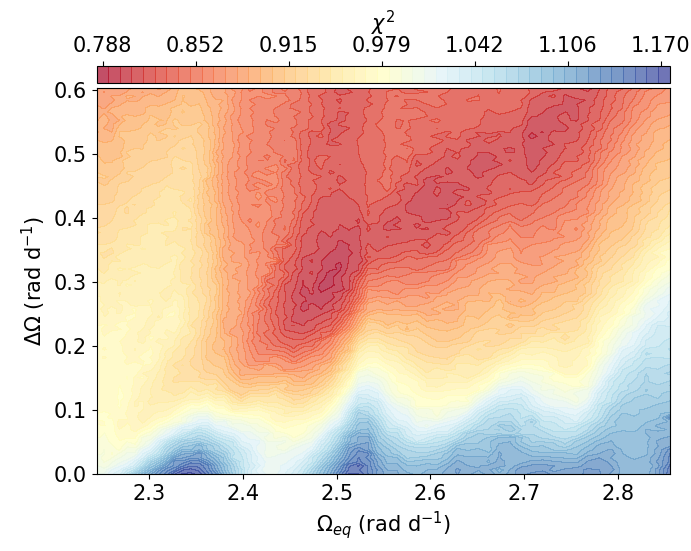}
   \caption{Two-dimensional grid search for the differential rotation parameters using Stokes \textit{I} data of EK Draconis.
   }
    \label{fig:drgridsearch}
\end{figure}

Although \citet{Waite2017} did not find a measurable surface differential rotation using LSD intensity profiles, we ran a grid search to determine the differential rotation. We assumed
\begin{equation}
\Omega(\lambda)=\Omega_{\rm eq}-\Delta\Omega\sin^2\lambda, 
\end{equation}
where $\Omega$ is the angular velocity at a given latitude $\lambda$, $\Omega_{\rm eq}$ is that at the equator, and $\Delta\Omega$ is the pole-equator difference. 
We searched for the pair ($\Omega_{\rm eq}$, $\Delta\Omega$) that minimises $\chi^{2}$. The resulting $\chi^2$ landscape is shown in Fig.~\ref{fig:drgridsearch}. The solutions did not converge to a global minimum, i.e., a reliable estimate for $\Delta\Omega$. Nevertheless, the deepest minimum lies within the range of strong shear rates estimated by \citet{Waite2017}, using Stokes V data (about 4 times solar surface shear). It is also consistent with the shear derived from the range of photometric periods measured by \citet{messina03}. Because our results do not provide a significant estimate for $\Delta\Omega$, we present the resulting sheared spot map only in the Appendix~\ref{sec:appenc}. We have found that differential rotation of this magnitude mainly affects the shape of the spots, in the sense that spots appear more sheared, owing to latitudinal differential rotation. Also, less spots are reconstructed at low latitudes. To evaluate the effect of such strong surface shear levels on the surface reconstructions, we performed additional simulations using the FEAT model (see Section~\ref{sec:models}), of which further details are given in Appendix \ref{sec:appenc}.

\subsection{Comparing contemporaneous images}
\label{ssec:compare}

    The most recent DI study of EK Dra was performed by J18, using very high resolution (R$\sim$230,000) spectral data, covering nine days. They obtained ten spectra between 3-11 April 2015, which is about 3.5 months earlier than our observations. This relatively short time interval between the observing runs provides an opportunity to compare the two maps. However, given the indications for strong differential rotation \citep[][and also our Fig.~\ref{fig:drgridsearch}]{Waite2017}, a detailed comparison of spots between the two maps is likely unreliable, especially in the  longitudinal direction. We emphasise that the following is a qualitative comparison between the overall spot distributions, in particular regarding the latitudinal zones where the spots are found.
    
    \begin{figure}
	\includegraphics[width=\columnwidth]{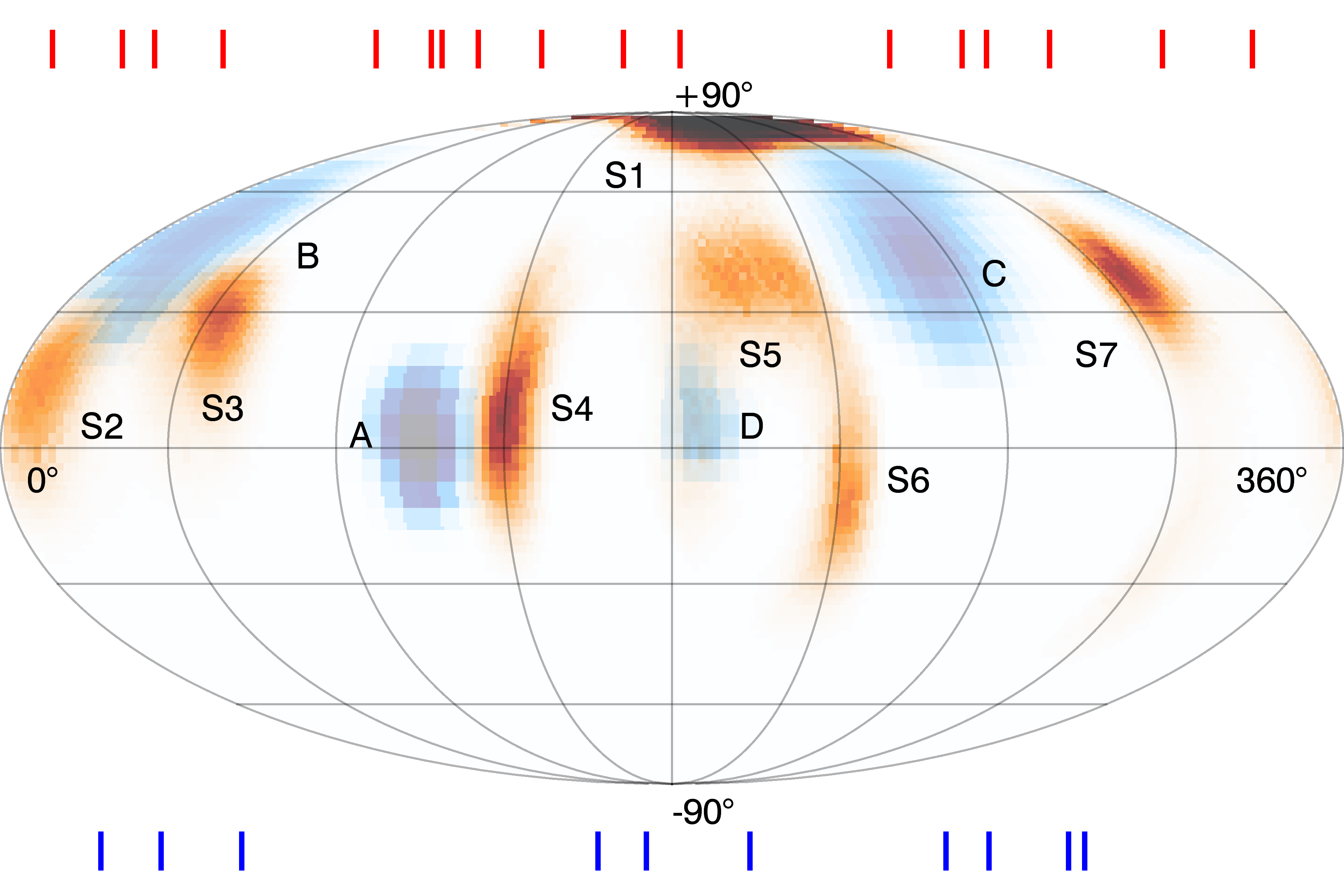}
   \caption{A comparison of the spot distribution of EK Dra obtained in this study (labelled as S1 to S5) and by \citet{Jarvinen2018} (labelled as A to D). The red and blue stripes show the distribution of spectral data obtained in this study and by \citet{Jarvinen2018} along stellar longitude.}
    \label{fig:ekdramw3}
\end{figure}

    We show in Fig. \ref{fig:ekdramw3} a comparison of the two Doppler images, based on the same ephemeris, so that the rotational phases are congruent in both cases for the equator. Our near-polar spot S1 has no counterpart in the J18 image. However, both images show spots at mid-latitudes ($30^\circ$ to $60^\circ$, B-C vs. S3, S5, S7) and across the equator (A, D vs. S2, S4, S6). We suggest a possible explanation for the cross-equatorial spots in Section~\ref{ssec:spots}. 
    
    The occurrence of a strong near-polar spot in our map indicates a substantial change in the large-scale distribution of activity within 3.5 months. Such rapid variation was visible also in the surface maps by \citet{Waite2017}, for December 2006, January and February 2007, as the high-latitude spot had disappeared within 3 months. In addition, their January 2008 map shows a predominant near-polar spot, which is very similar to the one in Fig.~\ref{fig:sphmap}. The difference with the surface map obtained by J18 may indicate that we have captured an episode of polar-spot formation. 







\section{Modelling the spot distribution}
\label{sec:models}
We used the Flux Emergence And Transport (hereafter FEAT) modelling platform developed by 
\citet{isik18}, to simulate the surface distribution of spotted regions on EK Dra. This platform 
consists of 
three parts: (1) a solar-like butterfly diagram of emergence latitudes, times, and sizes of 
spot groups through a single 11-year cycle; (2) the rise of magnetic flux loops from the base 
to the top of a solar-like convection zone model, which yields emergence latitudes and tilt 
angles; 
(3) surface transport of emerging magnetic flux in the form of bipolar magnetic regions 
(hereafter BMRs). 
The resulting time-dependent surface distribution of the signed magnetic field 
is then used to obtain the distribution of dark spots, using the simple approach taken by 
\citet{isik18}: 1-degree-squared pixels with a field strength above 187~G is considered as 
pertaining to a starspot.
This value was determined by matching the observed mean spot coverage 
on the Sun during a typical activity maximum, by taking into account pixels above this 
threshold in SFT snapshots. 

In this way, we confront the 
simulated FEAT maps with the Doppler image obtained in Sect.~\ref{sec:paramsdi}. 
In this paper we do not give a full description of the FEAT model, for which we refer the reader to 
\citet{isik18}. 
Before briefly describing the steps (1)-(3), we first assess the validity of adopting a solar 
internal structure model for EK Dra, when modelling the dynamics of flux tubes in step (2). 

\subsection{Convection zone geometry}
\label{ssec:mesa}
We compare the convection zone geometry 
of a solar internal structure model with a corresponding model for EK Dra, 
using Modules for Experiments in Stellar Astrophysics \citep[MESA][]{Paxton2011, Paxton2013, 
Paxton2015, Paxton2018, Paxton2019}. Taking the initial mass to be $1.04\,M_\odot$ 
(Sect.~\ref{sec:fundapars}, Table~\ref{tab:atmpar}) and using the default parameters defined 
in MESA (as of this writing) 
with $Z=0.02$, we modelled the pre-main-sequence evolution of EK Dra. We also produced a 
reference model for the Sun at an age of $4.69$~Gyr. 

\begin{figure}
\centering
\includegraphics[width=1.1\columnwidth]{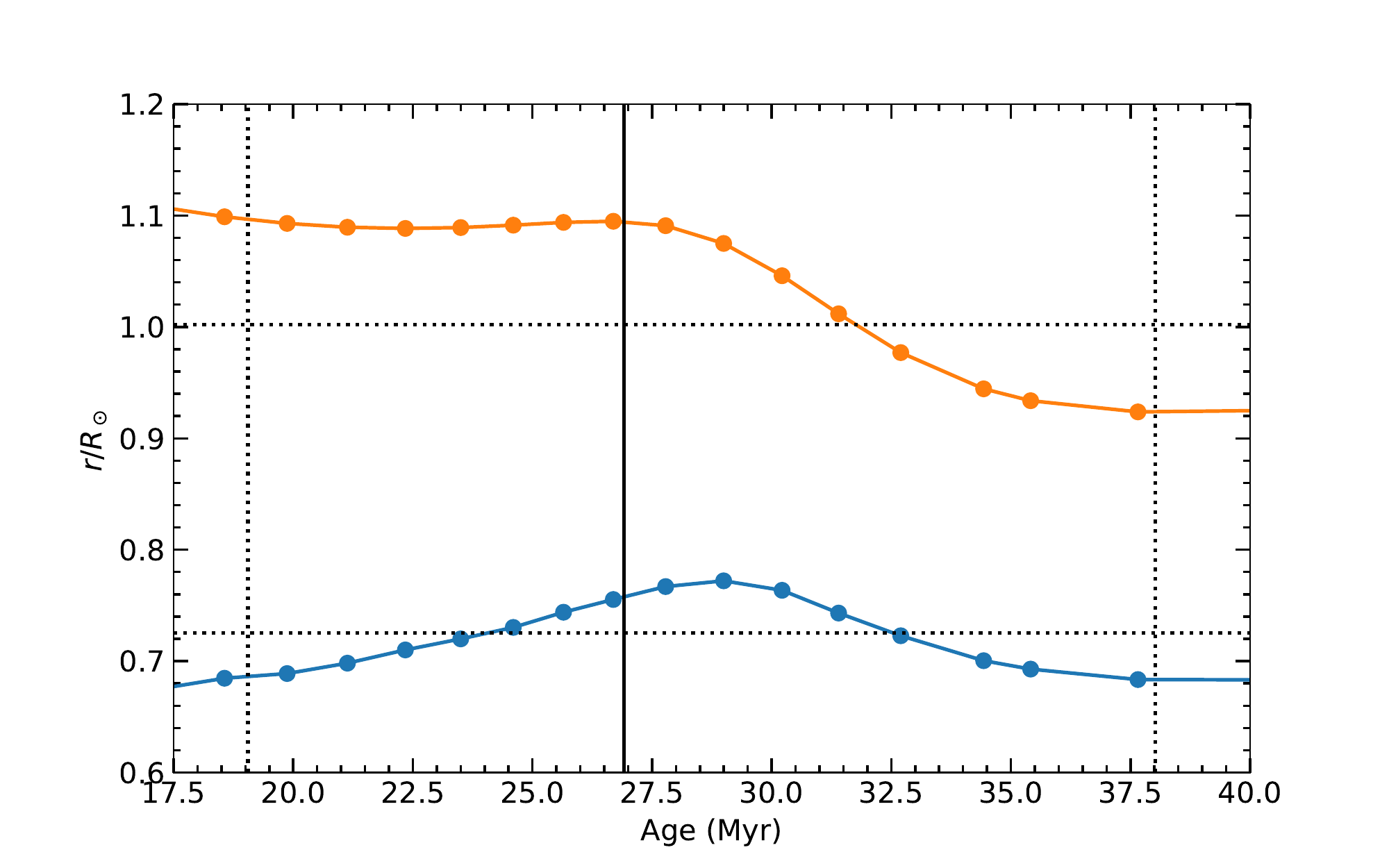}
\caption{Evolution of the bottom (blue) and the top (orange) radial locations of the convection 
zone in a stellar model with the mass of EK Dra. The radial locations are given as fractions 
of the radius of a solar model at an age of 4.69 Gyr, for which the horizontal dotted lines show 
the convection zone boundaries. The vertical lines show EK Dra's age (full line) and its error 
range (dotted lines).}
\label{fig:czevo}
\end{figure}

Figure~\ref{fig:czevo} shows the evolution of the base and the top locations of the convection 
zone from the MESA solution. The age and its error range indicated in the figure are taken from 
our estimate for EK Dra (Sect.~\ref{sec:fundapars}). 
It turns out that the convection zone of EK Dra must have a very similar aspect ratio to that 
of the Sun. This result is consistent with our finding that EK Dra and the Sun are very close 
to each other in the $T_{\rm eff}-\log g$ diagram (Sect.~\ref{sec:fundapars}, Fig.~\ref{fig:evol}). Considering the proximity of EK Dra to the Sun on the $T_{\rm eff}-\log g$ diagram along with 
Fig.~\ref{fig:czevo}, we suggest that EK Dra has a more 'Sun-like' internal stratification 
than that of a young $1\,M_\odot$ star at an age between 20-40 Myr! 
For this reason, we adopted the non-local mixing-length solar convection zone model  when modelling the emergence of magnetic flux tubes (Sect.~\ref{ssec:feat}).

\subsection{Emergence and transport of magnetic flux}
\label{ssec:feat}

We carried out numerical simulations using the FEAT module 
for the emergence of magnetic flux. As in \citet{isik18}, we set up a single 
activity cycle with a solar-like butterfly diagram at the bottom of the convection zone. 
The total number of emerging BMRs is scaled with the rotation rate with respect to that 
of the Sun, to account for the rotation-activity relation. EK Dra rotates at a rate of 
9.6 times the Sun, so we assume its BMR emergence frequency is also 9.6 times that of 
the Sun throughout its activity cycle, which we simply assumed to be 11 years long. The models 
presented here are close to the model considered by \citet{isik18}, for which the 
rotation rate and the activity level are 8 times that of the Sun. 

Taking into account the solar convection-zone stratification (Sect.~\ref{ssec:mesa}), 
we tracked the trajectories of magnetic flux tubes rising throughout the convection zone, for the 
rotation rate of EK Dra. For these computations we assumed that the internal rotation profile is the same as that of the Sun. The initial field strengths of the tubes were determined
using a linear stability diagram for flux rings at the bottom of the convection zone, 
calculated for the rotation rate of EK Dra, using also a 
solar-like internal differential rotation profile with the solar absolute rotational shear. 

\begin{figure*}
	\includegraphics[width=0.8\textwidth]{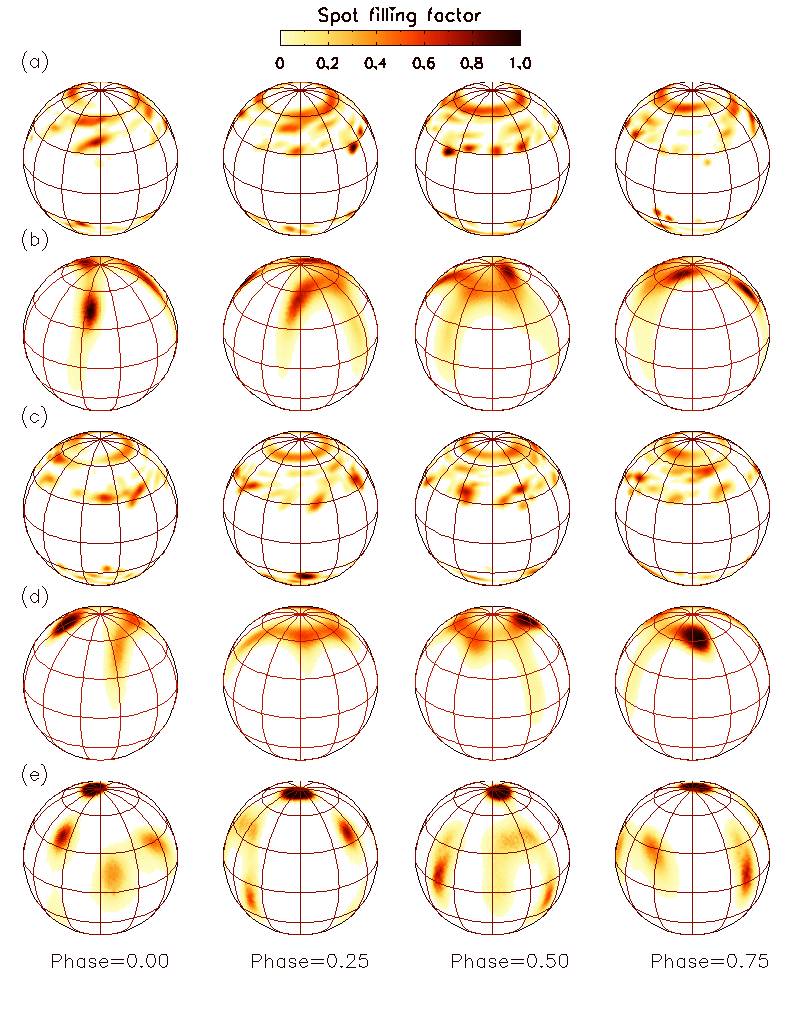}
   \caption{Comparison of FEAT simulations with the Doppler image of EK Dra. Panels (a) and (c) 
   are 15-day averaged simulations as seen on four rotational phases, (b) and (d) show their 
   corresponding synthetic Doppler images. Panel (e) shows the Doppler image (Fig.~\ref{fig:sphmap}) reconstructed from the observations.}
    \label{fig:ekdra_all}
\end{figure*}

As the final stage of FEAT, we carried out a surface flux transport (SFT) simulation covering 
11 years of a solar-like cycle, for which the emergence latitudes and tilt angles were taken 
from the flux-tube simulations described above. We assumed the same latitudinal 
rotation profile as on the Sun, 
so that the pole-equator lag is the same as on the Sun. 
Two examples of the resulting surface images are shown in 
Fig.~\ref{fig:ekdra_all}a,c, which are 3.5 months apart from each other, to evaluate possible extent of spot evolution within the time frame limited by our spectra and those of J18. To simulate the time spread of the spectra we used for DI, we averaged SFT snapshots with 1-day resolution over 15 days. The phase of the solar-like 11-year activity cycle shown here corresponds to about two years before the activity maximum. A near-polar spot region with a wide longitudinal extension is present, along with mid-latitude activity at smaller spatial scales. We note that each pixel on these images has an angular area of 1 degree-squared. The combined effect of mid-latitude emergence of strongly tilted BMRs, differential rotation, poleward meridional flow, and supergranular diffusion are responsible for the formation of the polar spot here 
\citep{schuessler1996,schrijver01,isik07,isik11,isik18}. 


\section{Discussion}
\label{sec:discuss}

\subsection{Chemical composition}
\label{ssec:chem}

The effective temperature ($5770\pm80$\,K) and surface gravity ($\log g=4.4\pm0.1$) we derived using various photometric and spectroscopic methods agree well with those derived by \citet{Strassmeier1998,Konig2005,Jarvinen2007,Jarvinen2018} within the uncertainty limits. The derived microturbulent velocity, $\xi=1.7\pm0.1$ km\,s$^{-1}$ agrees also well with that of \citet{Jarvinen2007}, which is 1.6 km\,s$^{-1}$, and it is slightly lower than that of \citet{Strassmeier1998}, which is 2.0 km\,s$^{-1}$. These atmospheric parameters are very similar to those of the Sun (Fig.~\ref{fig:evol}).

We found the elemental abundances of the atmosphere of EK Dra to be nearly solar for all studied elements within $\pm 0.21$ deviation, except for Li and Ba (Fig. \ref{fig:abundances}). The overabundance of [Li/H] = 2.18 indicates the youth of the star, i.e., $\tau \lesssim 50$\,Myr, and is consistent with the age we found from the $T_{\rm eff}-{\rm log}g$ diagram ($27^{+11}_{-8}$\,Myr). The Ba abundance was, however, most likely overestimated due to the assumption of depth-independent microturbulent velocity \citep[see][]{reddyandlambert17}. We estimate the corrected one as ${\rm [Ba/H]}\approx 0.18$. With this correction, the abundance pattern of EK Dra shows that Sr, Y, Ba, and Ce are marginally overabundant with respect to the Sun. The common trait of these four elements is that they are dominated by s-process nucleosynthesis for the solar system \citep[see Fig. 20 of ][]{wallerstein97}. The enrichment of these elements indicates that the progenitor low- and/or intermediate-mass AGB stars might have contributed the enrichment of EK Dra's birthplace somewhat more than in the case of the solar nebula. The small but systematic underabundances of Co, Ni, Cu, and Zn indicate that supernovae have affected the birthplace of EK Dra to a lesser extent, in comparison to the pre-solar environment. However, we note that all such apparent abundance deviations of trace elements can result from the assumptions we used for the abundance analysis: the assumption of LTE, neglect of the 3D structure of the stellar photosphere, omission of magnetic field, and depth-independent microturbulent velocity. A more detailed 3D non-LTE atmosphere modelling can allow a more accurate assessment of the deviations from solar abundances in EK Dra's atmosphere. 

The position of EK Dra on the $T_\mathrm{eff}$-log\,$g$ diagram suggests that the star is on the pre-main-sequence with a mass of $1.04\pm0.04$\,M$_\odot$, most-likely being a post-T Tauri star. \citet{Montes2001} have mentioned that EK Dra is a member of the Local Association, using all three components of its space velocity. \citet{lopez06} also suggested that the star belongs to the Local Association B4 subgroup, using kinematic, spectroscopic, and photometric criteria. Using the {\it Gaia} DR2 data, we derived the space velocity components as $U=-5.98$ km\,s$^{-1}$, $V=-30.15$ km\,s$^{-1}$, and $W=-3.56$ km\,s$^{-1}$. In the $UV$ velocity plane, components match those typically measured in Pleiades cluster, subgroup B4, and AB Dor moving group. However, the estimated young age of the star is either smaller or at the lowest limit of these groups. This makes it difficult to clearly reveal which cluster/association/moving group the star belongs to. The W velocity component is also not close to any of those of these groups. The difference seen in the W component may be due to the radial velocity of the EK Dra being disturbed by the secondary component.

\subsection{Spot distribution: observations vs. models}
\label{ssec:spots}

We now confront the FEAT simulations presented in Fig.~\ref{fig:ekdra_all}a,c with the spot distribution 
inferred by Doppler imaging (Fig.~\ref{fig:ekdra_all}e). To make them more compatible, we first rescaled 
the unsigned magnetic flux density (filtered between 187 and 800 G) to spot filling factor, so that any 
pixel with $|B|<187$~G has no contribution, and the pixel with $|B|\geqslant 800$~G has a filling factor of 
unity. Next, we generated synthetic LSD profiles at the same rotational phases as for our spectra. Using 
these profiles we produced two Doppler images corresponding to the two simulated epochs 3.5 months apart. 
This time interval was chosen to comply with the time difference between the spectra used by J18 and 
in this study (see Sect.~\ref{ssec:compare}, Fig.~\ref{fig:ekdramw3}). 

In the simulated DIs (Fig.~\ref{fig:ekdra_all}b,d), it is evident that only relatively large structures were reconstructed. The main reason for this is that smaller individual spot groups do not distort the line profiles sufficiently to have detectable modulation. Another reason 
is that the maximum-entropy algorithm is sensitive to the largest gradients in the input image. There is 
a qualitative parallelism between the observed image and the simulated one in Fig.~\ref{fig:ekdra_all}b, 
in that there is a strong near-polar and a mid-latitude (40-45$^\circ$ latitude) feature of comparable 
contrast. There are no spots in the latitude range $0^\circ-30^\circ$ in the simulated Doppler images, 
unlike the observed DI. 
This is related to the relatively high lowest latitude, which is determined by the Coriolis effect 
imparted on rising flux tubes. For future modelling, we propose that a latitudinal scatter in emergence latitudes 
can be included in the model. This effect is already thought to occur in the Sun, owing to 
convective buffeting of rising tubes. 
An alternative (and perhaps co-existing) mechanism is that  
spot-producing flux tubes can form at any depth throughout the convection zone. Owing to a much shorter rising distance compared to that for the base of the convection zone, such tubes can be responsible for low-latitude spots. 
In some 3D magnetoconvection simulations, such toroidal flux systems are seen to self-organise in the midst of the convection zone \citep[e.g.][]{nelson13,nelson14}. These two effects, namely the latitudinal scatter and a more extended depth range for flux-tube formation, can be considered in the future as further improvements of the FEAT model. 

In the simulations (compare Fig.~\ref{fig:ekdra_all}a with b and c with d), the features which are extended over latitude down to the equator are 
signatures of mid-latitude spots in the southern hemisphere, which become prominent in some rotational 
phases. There are similar elongated features on the observed image, visible in all rotational phases, 
which can also be related to opposite-hemisphere spots. 
A prominent example is the spot that is highly extended to the southern hemisphere, centred at a negative latitude, near the rotational phase 0.35. 

On the second simulated image (Fig.~\ref{fig:ekdra_all}d), the similarity with the observed image is 
somewhat less, because the mean latitude of activity at this epoch (in the simulation) is closer to the 
pole, owing to poleward diffusion of highly tilted BMRs. For this reason, the strong southern feature in 
phase 0.25 in the input image does not lead to a large equatorward extension on the DI, but rather 
intensifies the high-latitude northern feature. 

The SFT models in Fig.~\ref{fig:ekdra_all}a,c show that within 3.5 months the spot configuration can 
change significantly, owing to a high emergence frequency of spots and the ongoing surface transport 
of magnetic flux. The resulting synthetic Doppler images (Fig.~\ref{fig:ekdra_all}b,d) do not show 
much spatial correlation either. 
This means that any spatial correlation between the two Doppler images in Fig.~\ref{fig:ekdramw3} (especially in longitude) can be coincidental. 


\section{Conclusions}
\label{sec:conc}

Based on the 23 elements analysed, we conclude that the chemical composition of EK Dra's atmosphere is very similar to that of the Sun, within $\pm 0.21$ deviation, except for Li (Fig. \ref{fig:abundances}). The excess of lithium is consistent with the age of the star, which we estimated to be $27^{+11}_{-8}$\,Myr using isochrones on the $T_{\rm eff}-{\rm log}g$ diagram. This supports previous findings that EK Dra is a pre-main-sequence star in the post-T Tauri phase of its evolution. Its position on the HR diagram is so close to that of the Sun, that this 1.04-solar-mass star is closer to the Sun in terms of its internal structure than for a 1-solar-mass star of the same pre-main-sequence age. 

Motivated by the first comparison of Doppler imaging and FEAT models, we conclude that the presence of near-polar and mid-latitude spots on Doppler images can be produced by a solar-type wave of activity in the convection zone, mediated by rotational effects on rising flux tubes \citep{ss92,isik18}. However, the presence of near-equatorial spots are not fully explained by this model. We showed that at least those DI features with large latitudinal elongation can be contaminated by southern-hemisphere activity, which can be poorly reconstructed with the DI technique. The more compact and round features can be related with actual low-latitude emergence. In this case, convective buffeting of rising tubes or a radially extended generation of toroidal flux (i.e., a distributed dynamo) can be responsible. We suggest that when such low-latitude emergences are much less frequent than for mid-latitude emergences, they would be preferably detected with DI, which is sensitive to longitudinal gradients and works more precisely for lower latitudes, owing to the larger extent in radial velocities. Forward modelling of Doppler images with physical models
\citep[e.g.,][who modelled ZDI images]{lehmann19} is promising for a deeper understanding of the observational manifestations of stellar activity patterns. 

\section*{Acknowledgements}
We would like to thank S.P. J\"arvinen for providing the data used in Fig~\ref{fig:ekdramw3},
Oleg Kochukov (Uppsala University) for his useful comments concerning the effect of the magnetic field on the derived abundances, and the referee, Pascal Petit, whose critical comments led to substantial improvement of the manuscript.  HV\c{S} acknowledges the support by the Scientific and Technological Research Council of Turkey (T\"{U}B\.{I}TAK) through the project 1001 - 115F033. DM acknowledges support by the Spanish Ministry of Economy and Competitiveness (MINECO) from project AYA2016-79425-C3-1-P. This work has made use of data from the European Space Agency (ESA) mission. This work has been partially supported by the BK21 plus program through the National Research Foundation (NRF) funded by the Ministry of Education of Korea.

    {\it Gaia} (\url{https://www.cosmos.esa.int/gaia}) data were processed by the {\it Gaia}
    Data Processing and Analysis Consortium (DPAC,
    \url{https://www.cosmos.esa.int/web/gaia/dpac/consortium}). Funding for the DPAC
    has been provided by national institutions, in particular the institutions
    participating in the {\it Gaia} Multilateral Agreement.




\bibliographystyle{mnras}
\bibliography{senavcietal_2020_final.bib} 




\appendix

\section{axial inclination search}
\label{sec:appena}

\begin{figure}
	\includegraphics[width=\columnwidth]{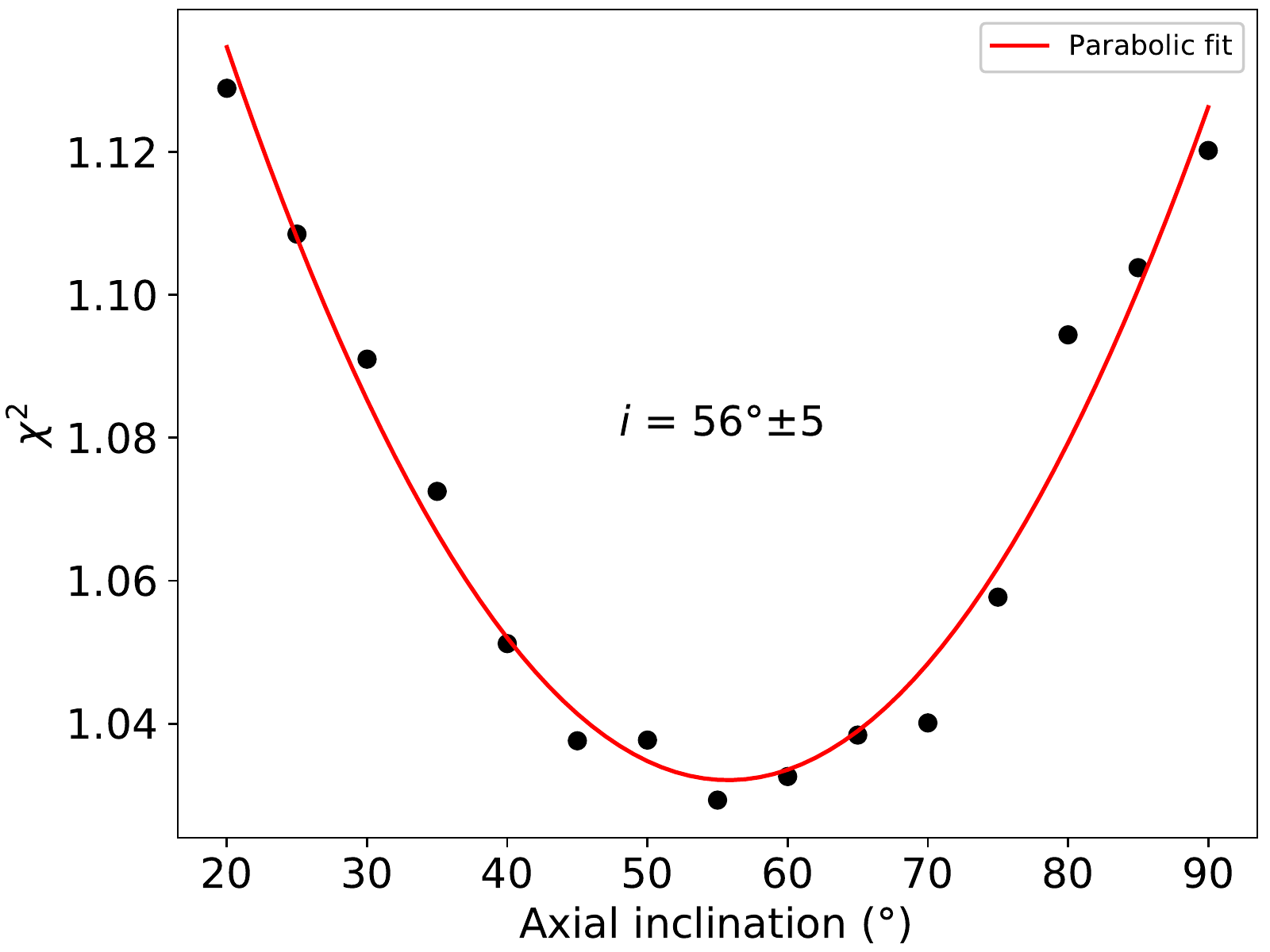}
   \caption{Results from a grid-search for the optimal axial inclination based on our data.}
    \label{fig:incsearch}
\end{figure}

We carried out a grid search for the axial inclination parameter using the Calar-Alto dataset. The resulting variation of $\chi^2$ is shown in Fig.~\ref{fig:incsearch}. We obtained the best axial inclination value as $56^\circ~\pm 5^\circ$ by fitting a parabola to $\chi^2$ values. Considering the formal error as $\pm 5^\circ$ and the systematic error to be higher, the resulting axial inclination value is consistent with the one obtained by J18, as well as the one by \citet{Waite2017}. It is also consistent with the other stellar parameters (e.g., the stellar radius) reported in the literature. We also note that replacing $\i=63^\circ$ adopted in this study (Table~\ref{tab:atmpar}) by $56^\circ$ has negligible impact on the reconstructed Doppler image.

\section{Optimisation of projected rotational velocity and the equivalent width parameter}
\label{sec:appenb}

\begin{figure}
	\includegraphics[width=\columnwidth]{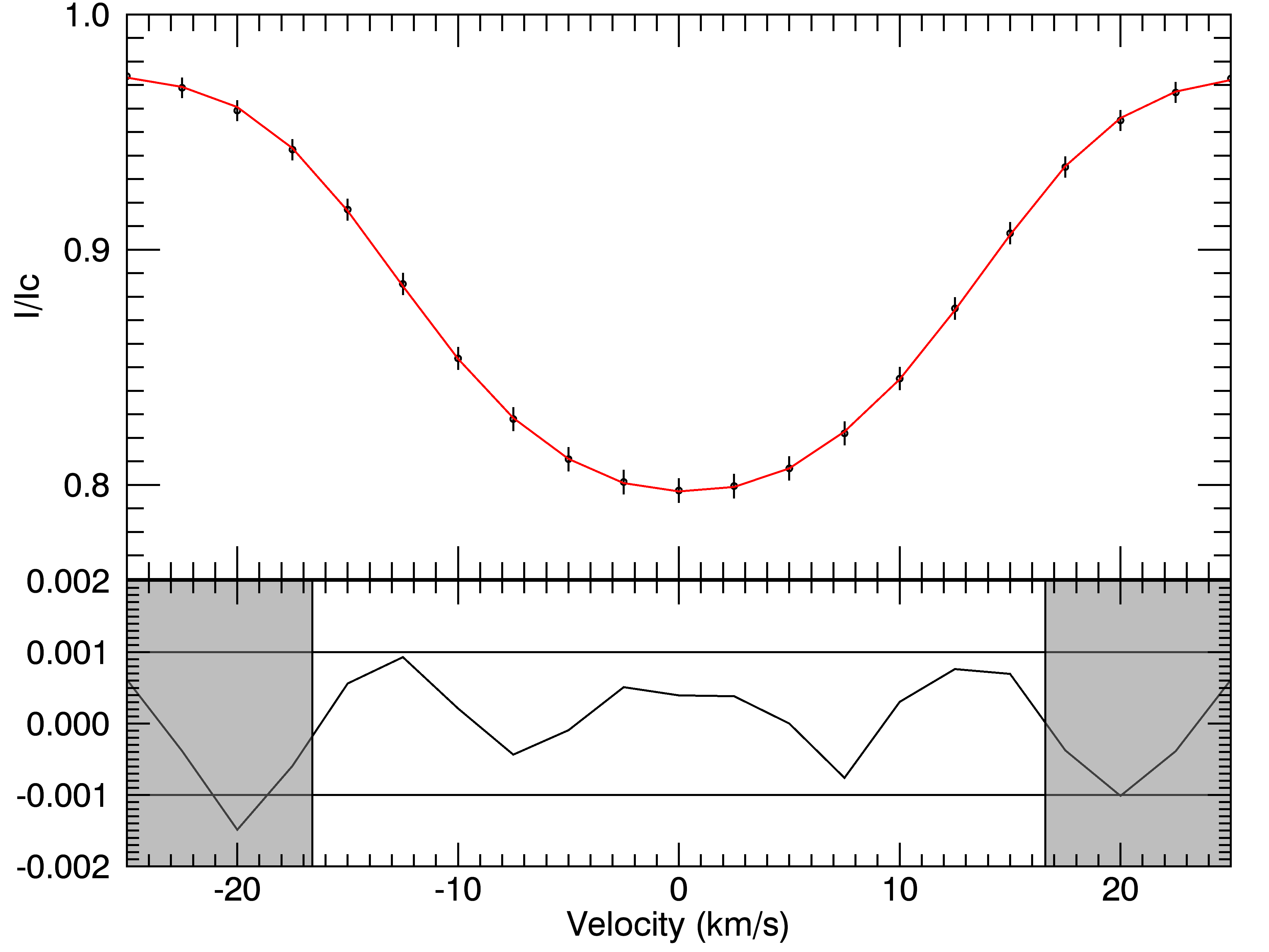}
   \caption{Phase-averaged LSD profile of EK Dra together with the error bars (top panel, black circles) obtained using the Calar-Alto data-set. Red solid line represents the maximum entropy regularised model obtained from the optimisation of the ${v \sin i}$ and the \textit{EW} parameters. Bottom panel shows the residuals from the model.}
    \label{fig:lsdfit_cang}
\end{figure}

We have optimised $v\sin i$ by requiring that the  phase-averaged residuals are minimised. This led to a $v\sin i$ of 16.7 km/s and a slightly higher \textit{EW} than the estimate from Fig.~\ref{fig:gridsearch}. This solution is not far from that of $\chi^2$ minimisation, for which a $v\sin i$ of 16.6~km/s was found. The phase-averaged LSD profile, the model using the optimised parameters and the corresponding residuals are shown in Fig.~\ref{fig:lsdfit_cang}. The morphology of the residuals is the same as that obtained for phase-ordered LSD profiles (lowest-right panel of Fig.~\ref{fig:lsdfits}). The residuals have an amplitude of about $\pm 0.001$, which is smaller than the residuals that were obtained by J18. The map reconstructed using the parameters achieved from the optimisation procedure is almost indifferent than the one imaged using parameters derived from $\chi^2$ minimisation.

\section{Simulations of strong differential rotation}
\label{sec:appenc}
There are multiple indications in the literature for the existence of strong solar-like differential rotation (DR) on EK Dra. \citet{messina03} used the observed range of photometric period from the observations covering $\sim$16 years and estimated an equator-to-pole difference in angular velocity, $\Delta\Omega$, of about 4 times the solar value. Based on 6 years of DR measurements with the ZDI technique, \citet[][see Table~6]{Waite2017} estimated a significantly large variance about the mean $\Delta\Omega=0.27$rad~s$^{-1}$, which is also about 4 times solar DR. Because it is not possible to reach a global $\chi^2$ minimum for $\Delta\Omega$ and $\Omega$ using only the Stokes I (intensity) profiles (Fig.~\ref{fig:drgridsearch}), we have presented the rigid-body-rotation solution in this study, as has been the standard approach in the field in such cases.

\begin{figure*}
	\includegraphics[width=.8\linewidth]{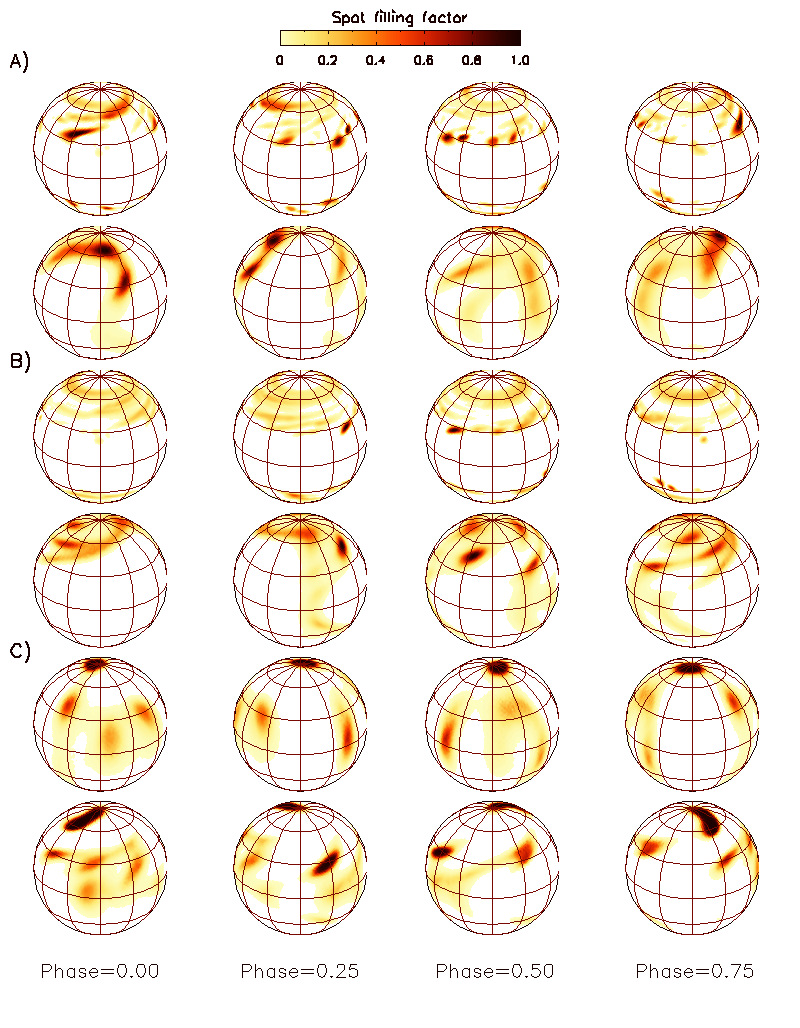}
   \caption{FEAT simulations and their reconstructions assuming 2 (panel A) and 4 (panel B) times the solar DR, in comparison to observed Doppler images of EK Dra for rigid-body and 4 times solar DR (panel C).}
    \label{fig:sftdr}
\end{figure*}

However, motivated by indications for strong DR on EK Dra, here we reconstructed DIs of SFT simulations that were run with 2 and 4 times solar DR ($\Delta\Omega$), in the same way as in Fig.~\ref{fig:ekdra_all}a-d, which were for solar DR strength. The results are shown in Fig.~\ref{fig:sftdr} (panels A and B), where we imposed the corresponding DR strength. The longitudinal and latitudinal elongations of the spot features (owing to increasingly strong shear) follow similar patterns in the reconstructed maps. In addition, there are low-latitude spotted regions in the reconstructed maps, though no such features exist in the input SFT maps. This is a combined consequence of the higher resolving power of the DI technique at lower latitudes and the coincidental phase encounters of southern-hemisphere spots in the FEAT models, as discussed in Section~\ref{ssec:spots}. In general, there is reasonable agreement between the input images and the reconstructions, indicating that sheared spot distributions can be well-reconstructed for up to 4 times solar DR. For comparison, we show in Fig.~\ref{fig:sftdr} (panel C) two Doppler images of EK Dra, reconstructed for rigid rotation and for 4 times solar DR, which is about the value corresponding to the minimum $\chi^2$ in Fig.~\ref{fig:drgridsearch}. In the sheared DI, there is considerably less spot coverage near the equator than in the rigid-body solution. In this sense, our FEAT models match more closely with the sheared DI of EK Dra. 

\bsp	
\label{lastpage}
\end{document}